\documentclass[%
 aip,
 amsmath,amssymb,
 reprint,%
floatfix]{revtex4-1}
\usepackage {CJKutf8}
\usepackage{graphicx}
\usepackage{dcolumn}
\usepackage{bm}
\usepackage{subfigure}

\usepackage[T1]{fontenc}
\usepackage{mathptmx}
\usepackage{multirow}\usepackage{etoolbox}

\graphicspath{{figures/}}
\usepackage[dvipsnames]{xcolor}
\usepackage{comment}

\makeatletter
\def\@email#1#2{%
 \endgroup
 \patchcmd{\titleblock@produce}
  {\frontmatter@RRAPformat}
  {\frontmatter@RRAPformat{\produce@RRAP{*#1\href{mailto:#2}{#2}}}\frontmatter@RRAPformat}
  {}{}
}%
\makeatother

\begin{document}
\begin{CJK*}{UTF8}{gbsn} 
\preprint{AIP/123-QED}

\title[Wang TY-PF-TAO]{Topologically assisted optimization for rotor design}
\author{Tianyu~Wang ({\CJKfamily{gbsn}王天宇})}%
\affiliation{School of Mechanical Engineering and Automation, Harbin Institute of Technology, Shenzhen 518055, China.}%

\author{Yannian~Yang ({\CJKfamily{gbsn}杨延年})}
\email{yangyn@scut.edu.cn, corresponding author}
\affiliation{Key Laboratory of Autonomous Systems and Networked Control, Ministry of Education, Unmanned Aerial Vehicle Systems Engineering Technology Research Center of Guangdong, School of Automation Science and Engineering, South China University of Technology, Guangzhou, 510640, China}

\author{Xuanwu~Chen ({\CJKfamily{gbsn}陈炫午})}
\affiliation{Institute of Aero Engine, Tsinghua University, Beijing 10084, China}

\author{Pengyu~Li ({\CJKfamily{gbsn}李鹏宇})}
\affiliation{Department of Mechanics and Aerospace Engineering, Southern University of Science and Technology, Shenzhen 518055, China}

\author{Angelo~Iollo}%
\affiliation{Institut de Mathématiques de Bordeaux, Université de Bordeaux and Memphis Team, Centre Inria de l'Université de Bordeaux, 33400 Talence, France}%

\author{Guy~Y.~Cornejo Maceda}%
\affiliation{School of Mechanical Engineering and Automation, Harbin Institute of Technology, Shenzhen 518055, China.}%
 
\author{Bernd~R.~Noack}%
\email{bernd.noack@hit.edu.cn, corresponding author}
\affiliation{School of Mechanical Engineering and Automation, Harbin Institute of Technology, Shenzhen 518055, China.}%

\date{\today}
\begin{abstract}
We develop and apply a novel shape optimization  exemplified
for a two-blade rotor with respect to the figure of merit ($FM$).
This topologically assisted optimization (TAO) contains two steps.
First a global evolutionary optimization is performed 
for the shape parameters and 
then a topological analysis reveals the local and global extrema 
of the objective function directly from the data. 
This non-dimensional objective function compares the achieved thrust with the required torque.
Rotor blades have a decisive contribution to the performance of quadcopters.
A two-blade rotor with pre-defined chord length distribution is chosen as the baseline model.
The simulation is performed in a moving reference frame 
with a $k-\omega$ turbulence model for the hovering condition.
The rotor shape is parameterized by the twist angle distribution.
The optimization of this distribution employs a genetic algorithm.
The local maxima are distilled from the data 
using a novel topological analysis 
inspired by discrete scalar-field topology.
We identify one global maximum to be located in the interior of the data 
and five further local maxima related to errors from non-converged simulations.
The interior location of the global optimum suggests that small improvements
can be gained from further optimization.
The local maxima have a small persistence, i.e., 
disappear under a small $\epsilon$ perturbation of the figure of merit values.
In other words, the data may be approximated by a smooth mono-modal surrogate model. 
Thus, the topological data analysis provides valuable insights
for optimization and surrogate modeling.
 
\end{abstract}

\maketitle
\end{CJK*}

\section{Introduction}
\label{sec:introduction}
\label{sec:intro}

The market of multicopters has grown rapidly in the past decade due to their relatively small dimensions in comparison to fixed wing aircraft and their maneuverability to fly in every direction, i.e., vertically and horizontally. 
These characteristics make them possible for penetrating in otherwise hard-to-reach areas, such as pipelines, bridges, and powerlines. 
Multicopters, similarly to helicopters, generate lift from the rotation of their rotor blades. 
In this type of drone, several rotors may be used and thus researchers have designed and fabricated different types of drones ranging from one to eighteen rotors~\citep{Hassanalian17}. 
To enhance endurance time and reliability of multicopters, the rotor design is one key factor.
Motivated by the recent high demand for multicopter drones, 
an aerodynamic optimization method of the rotor design is studied in this paper. 
In the case of urban applications, hover efficiency drives rotor or rotor design~\citep{Kwon13}. 
Therefore, performance optimization in hover condition is necessary. 

Historically,  the rotor performance was evaluated 
by the blade element momentum theory (BEMT)\cite{Weick1930book}. 
The optimal design has the inflow angle being constant along the radial direction (Betz condition) for minimum induced power, and each cross section airfoil is operated at or close to the maximum lift-to-drag ratio for minimum profile power~\citep{Leishman00}. 
This method has the input of rotor diameter, free stream speed, rotation speed, number of blades \citep{Wen2023e}, thrust (or power), radial distribution of lift coefficient and angle of attack, and the output is the radial distribution of chord length and twist angle~\citep{Hepperle10}. 
The Betz condition relies on the rigid vortex sheet to move backward undeformed, which is the case at high advance ratio conditions. 
Yet it is not valid at low advance ratio or static condition~\citep{Wald06}. 
Using simulations, 
the optimal design can be found through shape parameterization and parametric optimizers. 
A Genetic Algorithm (GA) and flow analysis based on computationally inexpensive blade element momentum theory (BEMT) were adopted for rotor optimizations~\citep{Kwon13, MacNeill18, Svorcan19}. 
The BEMT method were also used together with gradient method \citep{Vu13, Pacini21} and particle swarm optimizer~\citep{Wall12}.   
Myriad of other optimizers are available \cite{Weise2009book}.
The choice is based on the available data, the number of parameters 
and the assumed dependence of the objective function on the parameters.

When the rotor has complex geometry and is at high loading condition, BEMT was reported to be unreliable, as it does not take into account of the complex three-dimensional flow effect\citep{Montgomerie04}.
In addition, the induced velocity in BEMT is normalized by the free stream velocity~\citep{Adkins94}. This ambient flow vanishes  in the hovering condition of multicopters. Due to the limit of the BEMT, a significant improvement space is still possible for high fidelity tools (Computational Fluid Dynamics, CFD) together with optimizers. 
CFD simulation together with GA methods have been used to optimize the twist angle distribution for the helicopter blade during the hovering condition, and a surrogate model was used to reduce computation time~\citep{Leusink13}. 
For a boxprop optimization, 
a meta-model employing radial basis functions is used to interpolate on the obtained CFD results, which the GA uses to find optimal candidates along the obtained Pareto front~\citep{Patrao16, Patrao18}. Using the Wageningen B-series rotor polynomial expressions for thrust and torque coefficients, a preliminary ship propeller design was conducted together with Non-dominated Sorting Genetic Algorithm (NSGA)\citep{Xie11}.

A surrogate model was typically adopted for high fidelity methods during GA optimization.
The main reason was that this approach reduces the computational load. 
As the surrogate model adds uncertainty to the evaluation process, 
results directly from high fidelity methods are preferable. 
As high performance cluster becomes easily available and cheap, large number of rotor CFD simulations (order of hundred) can be finished in one week. 
Therefore, this research chooses the high fidelity method. 
Moreover, GA has the advantage for a global optimization, 
it becomes popular in rotor, propeller, and wind turbine designs~\citep{Burger07}. 
Therefore, the high fidelity method together with GA is adopted to optimize the rotor during hovering condition.

We observe that the results of optimization using genetic algorithm 
for the rotor have large uncertainty 
as converged simulations are computationally prohibitively expensive.
This problem is partially cured by augmenting the genetic algorithm
by adding ``anti-noise'' to the data for topological simplicity.
This novel procedure is called Topologically Assisted Optimization (TAO).
The TAO analysis includes two steps. 
In the first one, extrema are extracted from the data set inspired by discrete scalar-field topology ~\cite{Kasten2016am}.
The second step is simplifying the topology of the data manifold 
with an Elastic Response Model (ERM) 
inspired from elastic maps and nets~\citep{Gorban2008}. 
Following the very idea of discrete scalar-field topology, 
we emphasize that we neither use smoothing filters nor derivatives to minimize bias.
Hitherto, discrete scalar-field topology has, to the best of our knowledge,
only been applied to two and three-dimensional spaces 
due to computationally expensive combinatorial algorithms.
Our key innovation is to generalize main features  of this analysis to arbitrary numbers
of independent variables.

\begin{figure}
\includegraphics[width=0.45\textwidth]{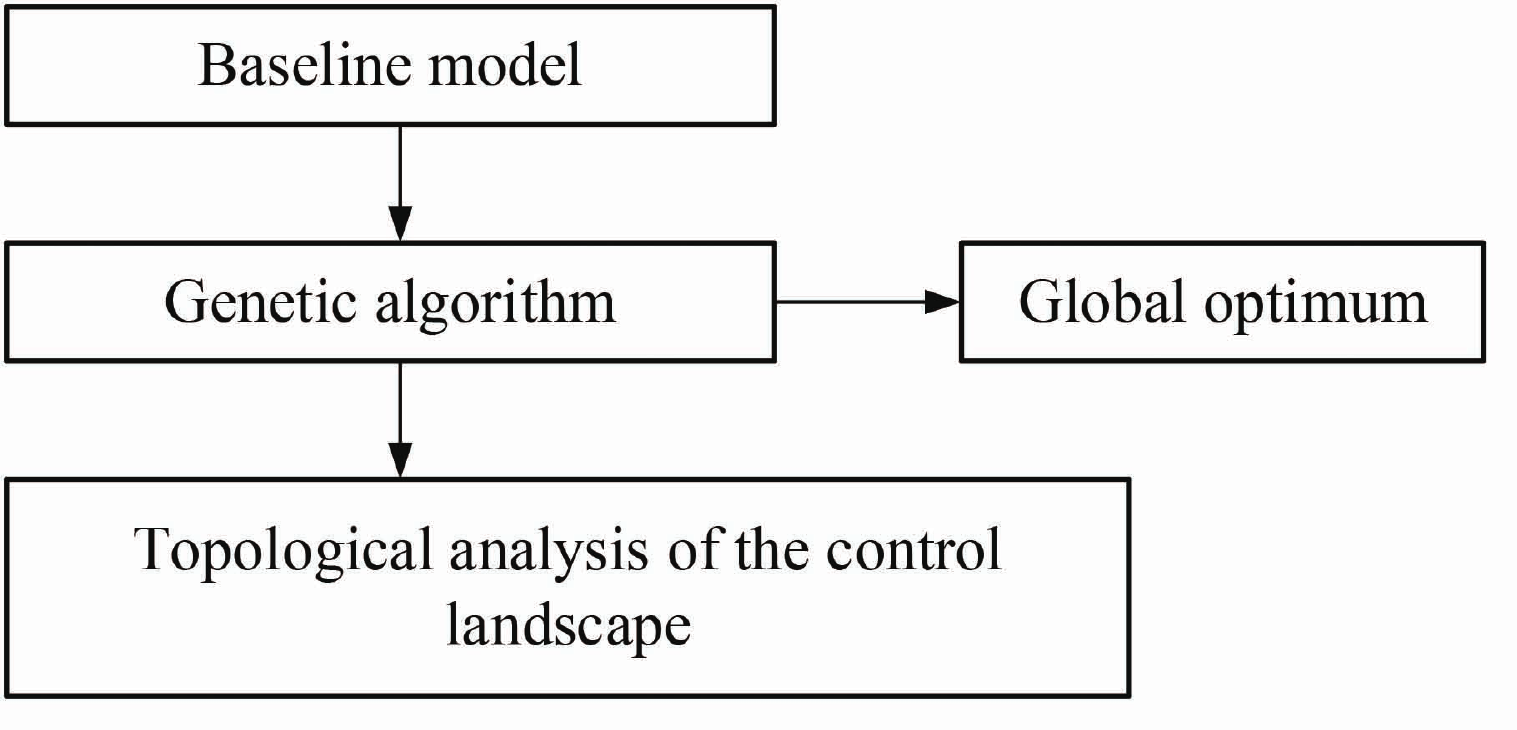}
\caption{Sketch for the topologically assisted optimization approach }
\label{fig:optimization_scheme}
\end{figure}

The optimization and analysis scheme is shown in Fig.\ref{fig:optimization_scheme}. 
Rotor design is described in Section \ref{sec:Propeller design}, 
and rotor optimization is conducted in Section \ref{sec:GA}. 
The neighborhood analysis is shown in Section \ref{sec:Neighbor analysis} 
as first insight from the data. 
TAO is explained in detail in Section \ref{sec:TAO}. 
The conclusions and outlook are provided in Section \ref{sec:Conclusion}.

\section{\label{sec:Propeller design}Propeller design }

\subsection{\label{sec:Parameterization of the shape}Parameterization of the shape}
\begin{figure}
\includegraphics[width=0.45\textwidth]{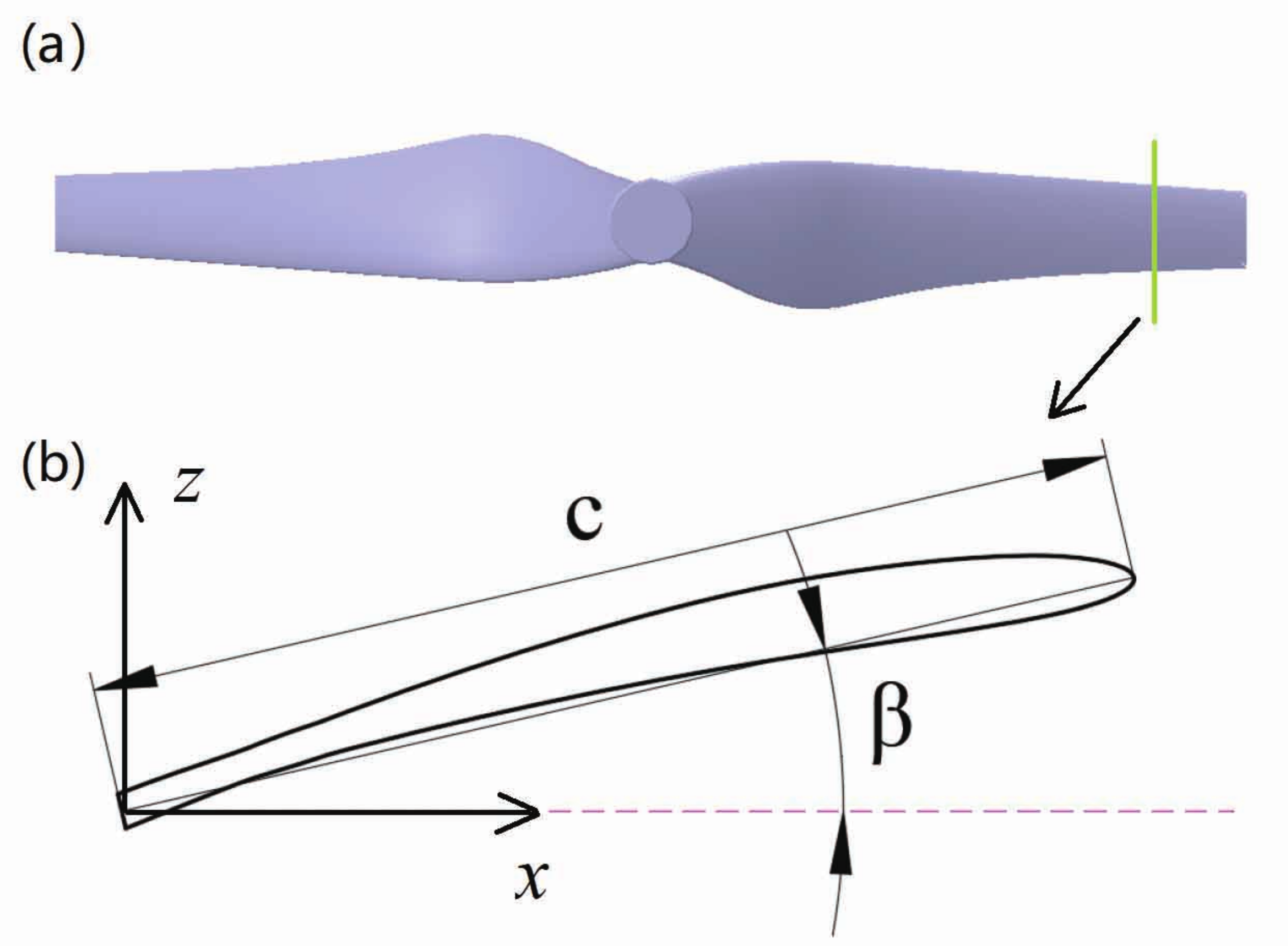}
\includegraphics[width=0.49\textwidth]{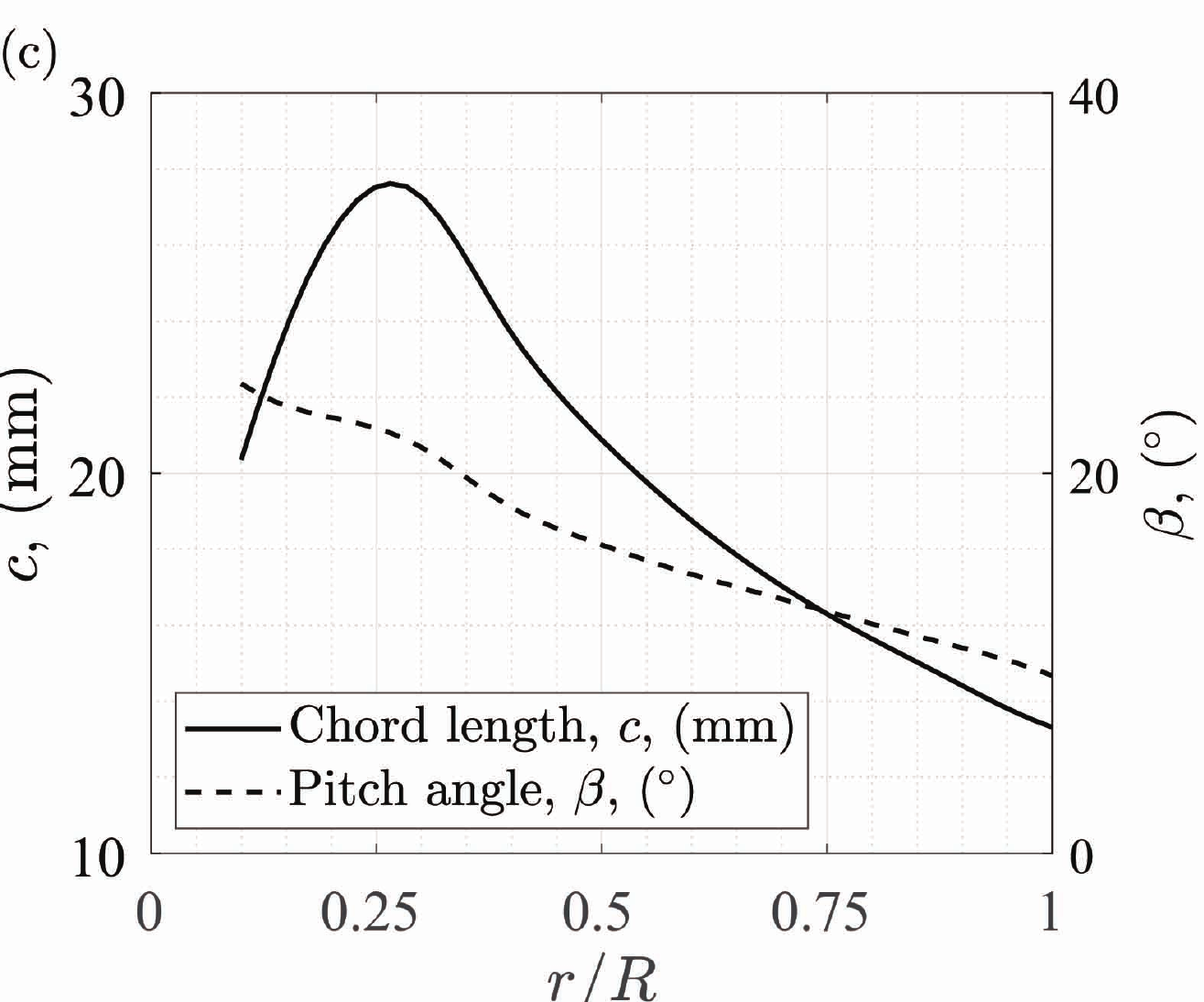}
\caption{(a) Top view of the rotor; (b) sectional view of the cross section and definition of the chord length and twist angle; (c) chord length and twist angle distribution as a function of the normalized radius $r/R$.
}
\label{fig:chord_pitch}
\end{figure}

The rotor of eHANG Ghost Drone 3.0 is taken as the baseline model, which is a light-weight UAV (about 800 grams) for video shooting and formation displays. The two-blade rotor has a diameter of 0.206m. The rotor is described by the profiles from the root to the tip, as well as the chord length and twist angle along the radial direction. The chord length and twist angle distributions of the model are shown in Fig.\ref{fig:chord_pitch} by the solid red and dashed blue curves, respectively. The airfoil chosen for the inner range ($r/R\le 0.1$) of the rotor is NACA 2416, and for the outer range ($r/R \ge 0.2$) is A18. The axis across approximately 45\% of the chord for different cross sections is aligned with the hub center. The radius of the rotor is $R$ = 0.103 m. More geometric data and experimental results of the baseline rotor can be found in our previous work\citep{Yang20}.

\subsection{\label{sec:FM}Figure of merit}
The power input for the rotor is the torque ($Q$) multiplied by the rotation speed ($\omega$), and the output is the thrust ($T$) multiplied by the speed in the axial direction ($u$). The far field velocity is zero during hovering condition, and thus the induced velocity at the rotor plane is adopted from the actuator disk model for output power calculation, which is $u = \sqrt{T/(2\rho A)}$ with $\rho$ the density of air. The actuator disk model is built from the momentum theory. A parameter named figure of merit ($FM$) is defined as

\begin{equation}
    \label{eq:FM}
    FM = \frac{Tu }{ Q \omega} =\frac{T^{3/2}/\sqrt{2\rho A}}{2 \pi Q (RPM/60)}
\end{equation}

\subsection {Computational domain, mesh, boundary conditions and solver}
\begin{figure}
\includegraphics[width=0.45\textwidth]{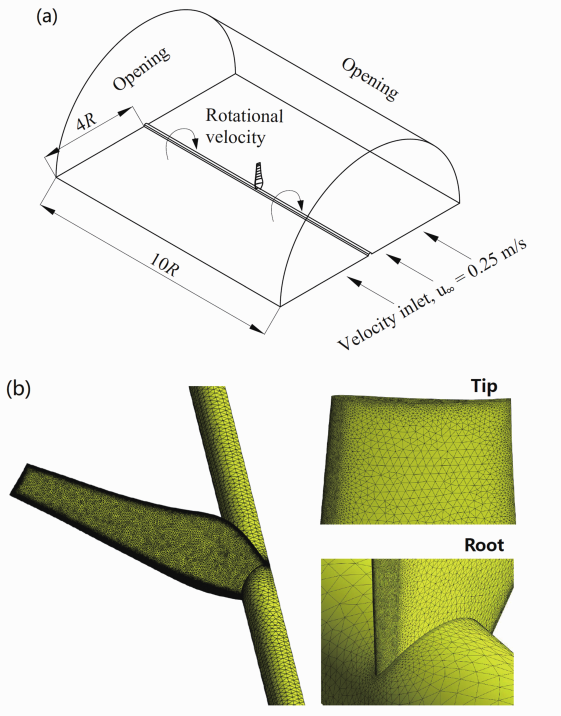}
\caption{(a) Computation domain and boundary conditions for the CFD simulations; (b) surface mesh on the blade and hub, as well as the zoom-in view on the blade tip and root.}
\label{fig:CFD}
\end{figure}

The technique of moving reference frame is adopted to simulate the rotor, which is operated at the hovering condition. The computation domain is a semi-cylinder with a radius of 4$R$, and a length of 10$R$ as shown in Fig.~\ref{fig:CFD} (a). Moving wall boundaries with no-slip wall conditions are set for the blades and hub. Unstructured mesh is employed for the whole domain. The mesh near the boundary layer has prism layers, with a $y$+ value of 1, and the rest is tetrahedral mesh. The mesh has a refinement in the region near the tips, leading edges and trailing edges of the rotor blades to capture high gradients of flow variables, which were assigned as separate surfaces while generating the geometry. The mesh on the blade and refinement regions are shown in Fig.~\ref{fig:CFD} (b).

\begin{figure}
\includegraphics[width=0.45\textwidth]{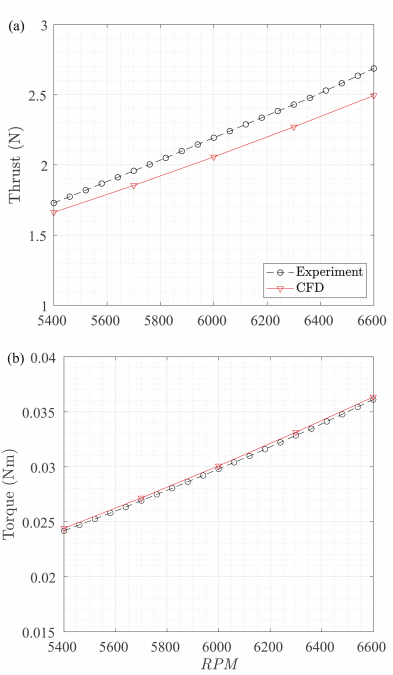}
\caption{CFD validation with experimental results. (a) Thrust versus $RPM$; (b) Torque versus $RPM$.
}
\label{fig:validation}
\end{figure}

The Reynolds number is $5.6 \times 10^4$ with the reference length of the chord length at the blade tip, and the reference velocity of the tangential velocity at the blade tip for the rotation speed of $RPM = 6000$. The turbulence intensity is 5\%. Several turbulence models
including the employed $k-\omega$ version have been validated for propeller simulations at moderate Reynolds number in our previous work. The time-averaged loading of the numerical results show good agreement with corresponding experimental results\citep{Yang17}. 
The numerical results have also been validated for this specific rotor by the experimental results at a sweep of $\it{RPM}$ herein.
The thrust and torque versus $\it{RPM}$ are shown in Fig.~\ref{fig:validation} (a) in the $\it{RPM}$ range of $[5400, 6600]$, centered around the design $\it{RPM}$ of 6000.
The numerical simulations are close to the experimental results.
The thrust value of CFD simulation at $RPM$ of 6000 is about 5.9\% lower 
than the balance measurement results, 
which may be attributed to the small inlet velocity and a ground effect. 
The torque value of CFD simulation at $\it{RPM}$ of 6000 is about 0.6\% larger than the balance measurement results, 
which may be attributed to the fully turbulent flow with a higher friction drag.

The inlet of computation domain is set as the velocity inlet boundary condition, which has a small velocity of 0.25 m/s to help the convergence.
The impact of the small inlet velocity is evaluated by checking the thrust value at 0.5, 0.25, and 0.1 m/s, and the thrust value has a small increasing trend as the inlet velocity decreases. The thrust value at 0.25 m/s is 0.6\% lower than that at 0.1 m/s, which can be viewed as infinitesimal and close to 0 m/s.  We also assume that the thrust increase caused by this low inflow velocity is the same for all models, and it does not deteriorate the optimization result. The outlet of the domain is set as a pressure outlet boundary condition with relative pressure of 0 Pa. The semi-cylindrical surface on the side is set as an open boundary condition. Periodic boundary condition is set for the two planes next to the hub. The hub is modified to extend from the inlet to the outlet to avoid the complex flow around the spinner and nacelle base, which has a benefit of saving computation resources. 

The mesh convergence was checked by the thrust value at the same $RPM$ of 6000 for three levels of node number, i.e., 1 million (coarse), 4 million (medium), and 9 million (fine). The thrust value on the rotor is 2.122, 2.147, and 2.149 N for 1, 4, and 9 million nodes, respectively. As the thrust value at the fine level of mesh only differs 0.1\% from that of the medium level, the medium mesh is selected for faster computation. 

Parallel computation is conducted on a cluster with commercial software of ANSYS CFX. Spatial discretization is based on a second order upwind scheme. Velocity and pressure coupling is achieved by the SIMPLE algorithm.  The CFD simulation of the rotor is conducted at the rotation speed of $\it{RPM} = $ 6000 for all models. 

\subsection{\label{Convergence problem}Convergence problem }
The convergence of the simulation is quantified by monitoring the thrust and torque forces on one blade, which represent the output and input of the rotor. The thrust value convergence history is shown during the 30,000 iterations. It can be observed that the thrust value converges around 25,000 iterations, which costs around 2600 core hours. As the genetic algorithms need to evaluate around 200 individuals in this study, the current iteration numbers for a full convergence are not affordable. Therefore, a compromise is made in this study by choosing a pre-converged data point, which is 900 iterations (80 core hours of computation)  for all individuals. Indeed, such pre-converged data will add significant uncertainty to the individual designs, and this negative impact is alleviated by our persistency method. 

\begin{figure}
\includegraphics[width=0.45\textwidth]{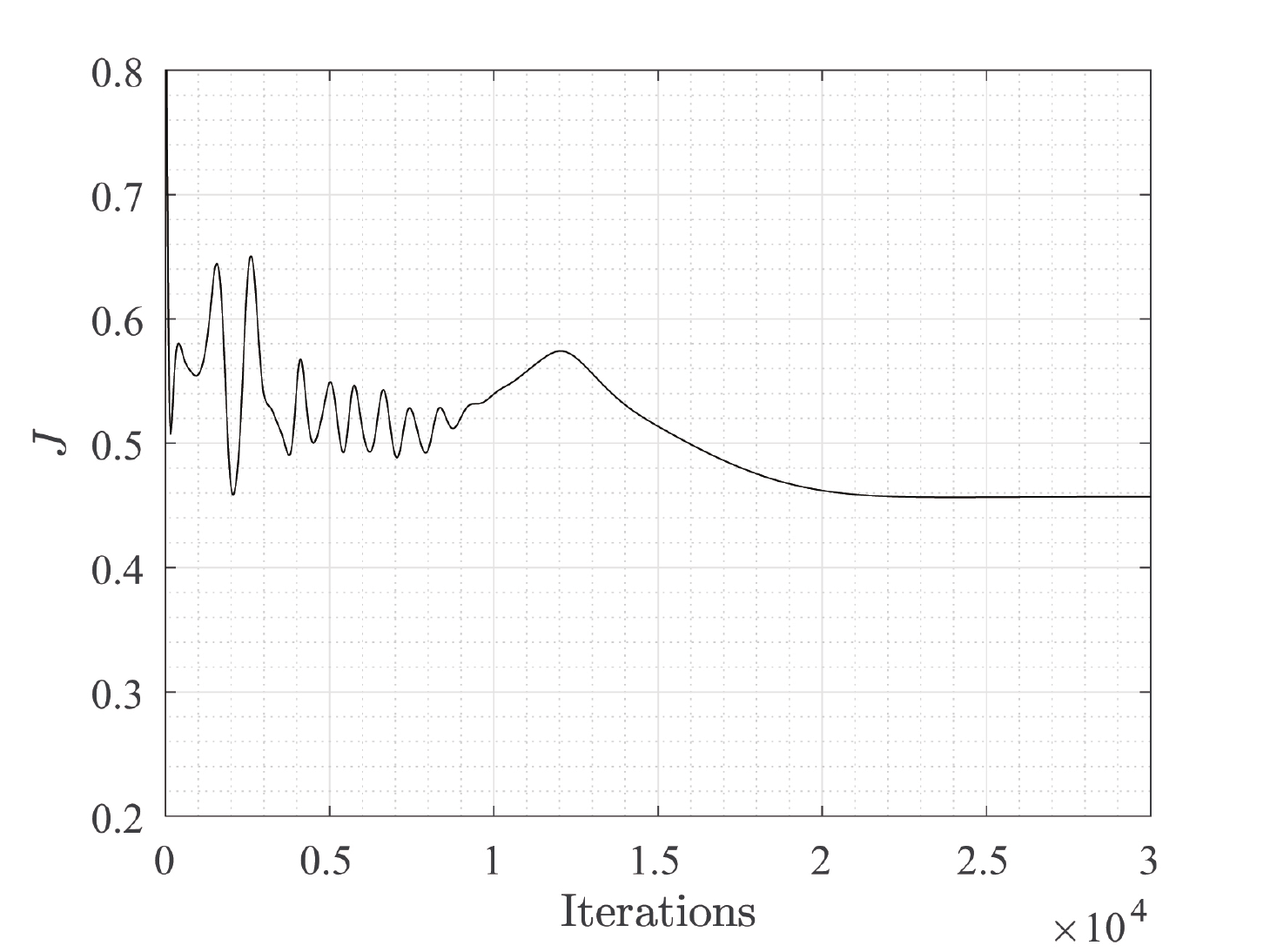}
\caption{Convergence history of $FM$ value of the baseline rotor.}
\label{fig:geo_par}
\end{figure}

\section{\label{sec:GA} Optimization with a genetic algorithm }

\subsection {\label{sec:Optimization setting}Optimization setting for a genetic algorithm}

For real applications, a prescribed chord distribution (planform) based on an exist model is common and advantageous. This way can facilitate manufacturing, motor compatibility, and customer acceptance. An optimization from a known chord distribution is called inverse design problem, and it is widely observed in industry~\citep{Hepperle10, Traub17,Xia22}. The chord length constrain limit for parameterization based on an exist model is also observed~\citep{Leusink13}. In order to upgrade an existing multicopter rotor, the chord length distribution is fixed in this research, and only the twist angle distribution is set as the input parameters. Observing the twist angle distribution of most rotors, it generally has a smooth decreasing trend from the root to the tip. Therefore, a third order polynomials are approximated as the fitting curve, which is found to well describe the current rotor. At the end, the twist angles at four cross sections, i.e., $r/R = 0, 0.4, 0.7$, and 0.95,  are chosen as the optimization parameters  

The optimization platform is based on the NSGA-II (Non-dominated Sorting Genetic Algorithm) algorithm~\citep{Deb02}, which is widely adopted for multi-objective optimization. The NSGA-II method replaces the non-dominated sorting to the magnitude sorting for a single objective optimization, which is the case in this study. The initial population of blade designs using the random sampling method, and CFD is conducted to evaluate the values of the objective functions ($FM$). First, individuals are ranked according to their objective values of $FM$. Then, a new generation of child individuals is created by crossover of the most interesting parent individuals. These children are consequently mutated and then evaluated by the simulation tool to obtain their objective values.

\subsection{\label{sec:Existing data}Existing data}

\begin{figure}
\includegraphics[width=0.45\textwidth]{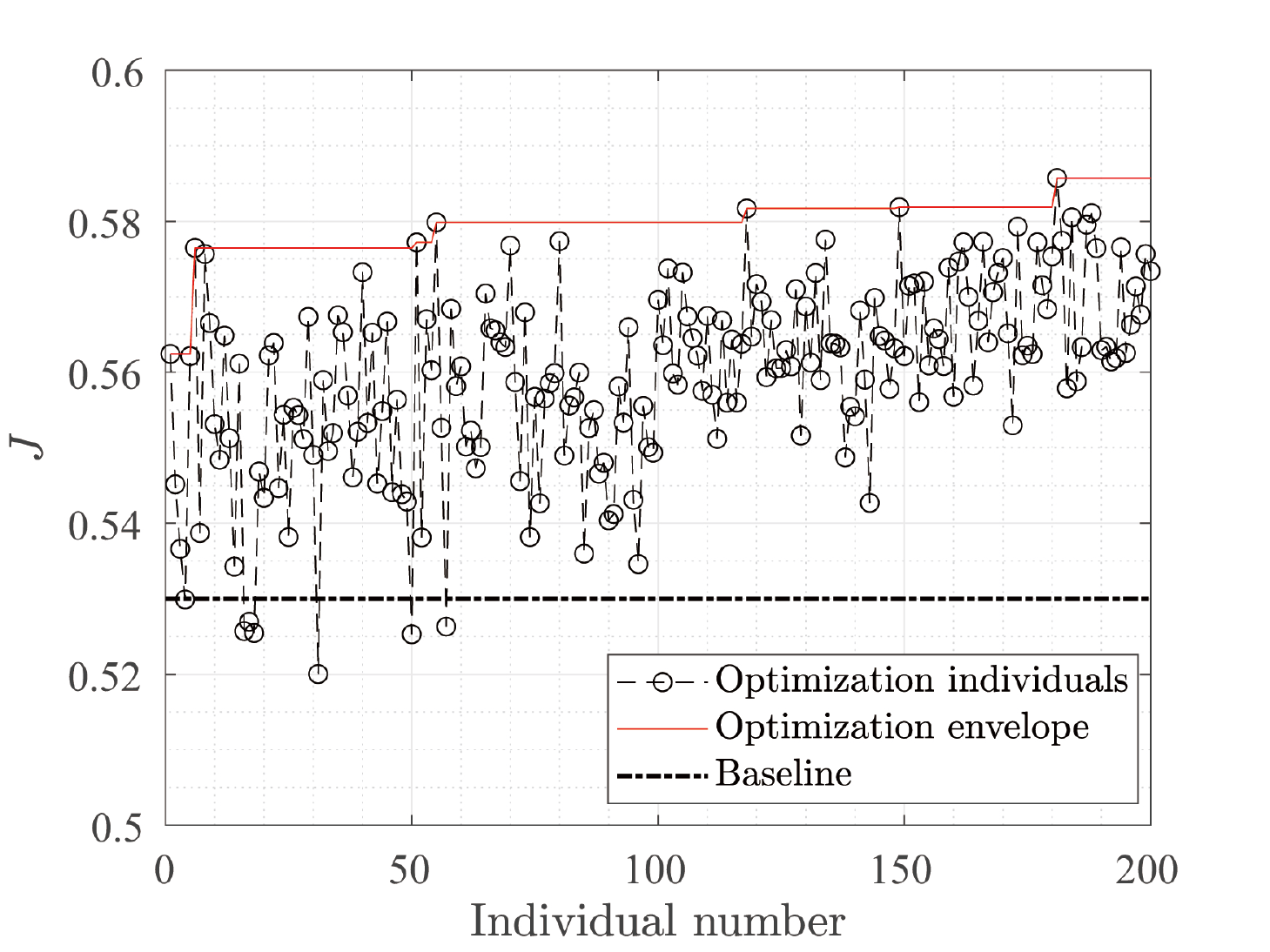}
\caption{Optimization process of genetic algorithm.
The red curve indicates the evolution of the best individual.}
\label{fig:opt}
\end{figure}

The results of the rotor optimization are shown in Fig.\ref{fig:opt}, which shows the 200 CFD cases evaluated for optimization, the maximum value envelope during the optimization process, and the performance of the baseline rotor. In comparison to the baseline rotor, an improvement of approximately 9.5\% in efficiency has been observed.

\section{\label{sec:Neighbor analysis} Neighborhood analysis of rotor data set }

In this section, we extract the local and global maxima of the data based on the analysis of the neighbors distribution of each point, i.e., finding the neighboring points and their relationship in terms of $FM$.

Let $\bm{b}^m \in \mathbb{R}^N$ be one set of parameters designating one individual, i.e., one rotor configuration.
For the rotor data $N = 4$, correspond to the twist angles at four cross sections employed for the optimization.
We note $J^m$ as the $FM$ associated to point $\bm{b}^m$; $J^m$ is the figure of merit $FM$ defined in Section~\ref{sec:FM}.
Let $B := \left \{  \bm{b}^m\right \}_{m=0}^{M-1} $ be the set of all data points.
$B$ is cleaned such that points with several evaluations, i.e., several figures of merit, are replaced with one copy whose $FM$ is the mean value of all the evaluations.
After cleaning the data, the number of points is $M=188$.

The extraction of the neighboring points is based on the distance between one point and the others.
Let $||\bm{b}^m - \bm{b}^n||$ be the Euclidean distance between two points $\bm{b}^m$ and $\bm{b}^n$.
Let $\mathcal{B}_L(\bm{b}^m)= \{\bm{b}^{k^m_1}, \bm{b}^{k^m_2}, \ldots, \bm{b}^{k^m_{L}}\}$ be the $L$-neighborhood of the data point $\bm{b}^m$ such that 
$|| \bm{b}^{k^m_1} - \bm{b}^m || \leq || \bm{b}^{k^m_2} - \bm{b}^m || \leq \ldots \leq || \bm{b}^{k^m_{L}} - \bm{b}^m ||$.
In the following, the superscript $k$ employed for a point designating a neighbor.

\begin{figure}
\centering
\includegraphics[width=0.50\textwidth]{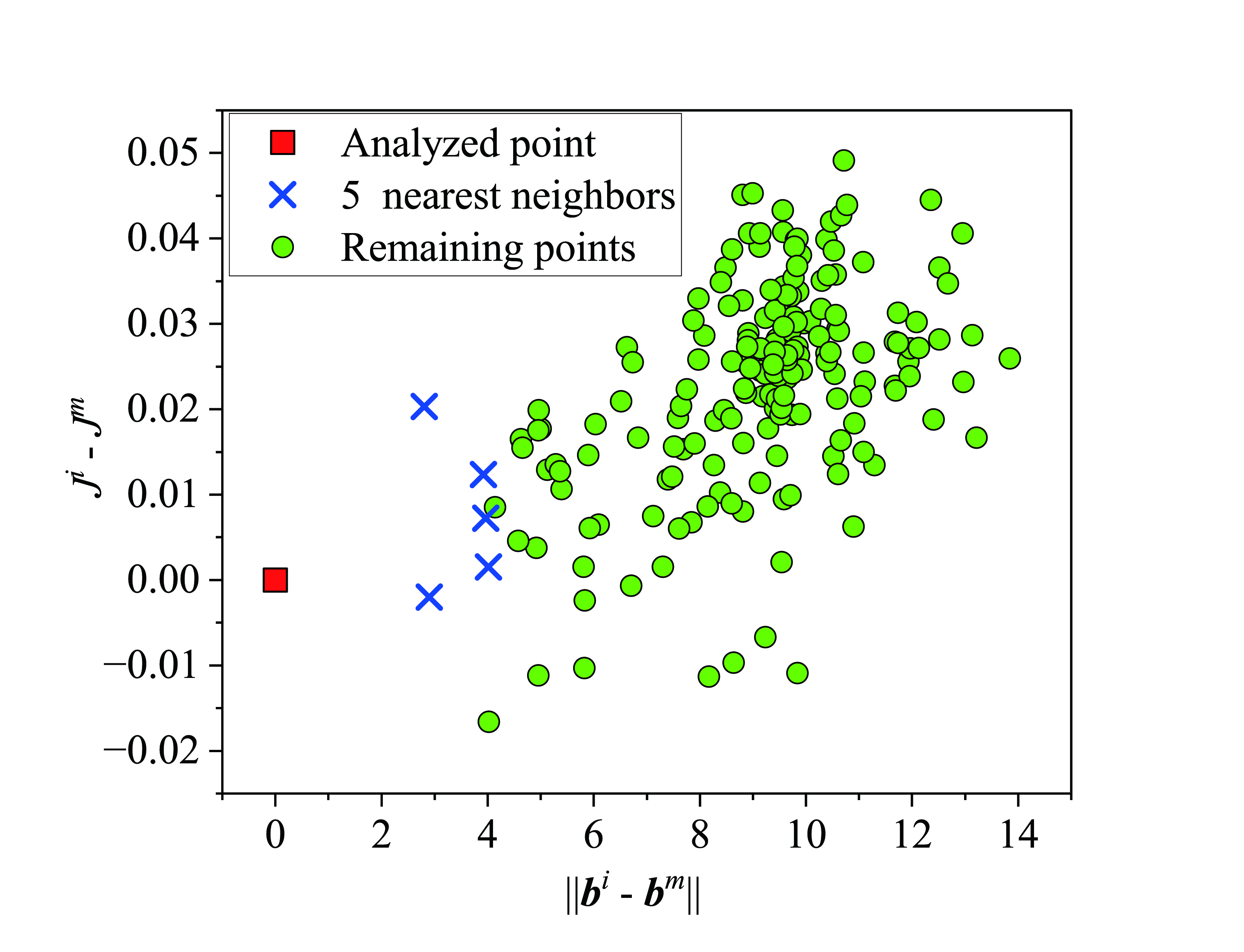}
\caption{\label{fig:randpoint}
$FM$ discrepancy versus distance between a random data point $\bm{b}^m$ (red square) and its 5 nearest neighbors (blue crosses) and the remaining data points (green dots).
The ID in the data set of this random point is 2.}

\end{figure}
Fig.~\ref{fig:randpoint} depicts a random point $\bm{b}^m$ (ID = 2) and its 5 neighbors versus their $FM$ discrepancy.
The distance between $\bm{b}^m$ and its nearest neighbor is 2.805, around $19\%$ of the data set diameter $D$, i.e., the maximum distance between two points of the data set.
The point $\bm{b}^m$ (ID = 2) is not a maximum of the data set as one of its 5 neighbors has a higher $FM$.

\begin{figure}
\centering
\includegraphics[width=1\linewidth]{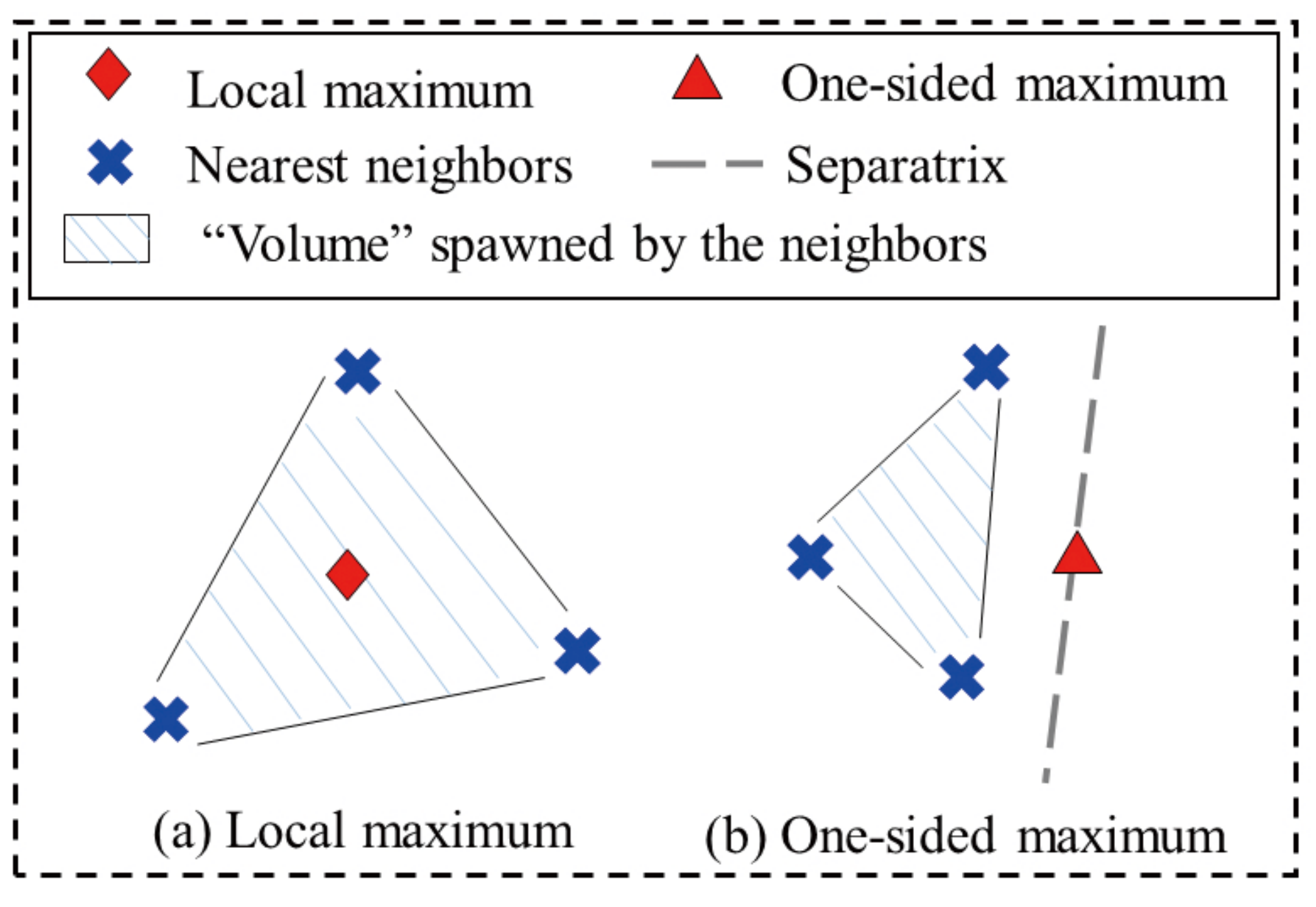}
\caption{\label{fig:ConditionC}
Scenario of the neighbors distribution of (a) local maximum (red diamond) and (b) one-sided maximum (red triangle).
}
\end{figure}

In this discrete data framework, a point $\bm{b}^m$ with $FM$ is a local maximum, see Fig.~\ref{fig:ConditionC}(a), if there is a $L$-neighborhood $\mathcal{B}_L(\bm{b}^m)$ with following properties:
\begin{itemize}
    \item \textbf{Property 1 (maximality):} $J^{k^m_i} \le J^m$ for $i=1,\ldots, L$ where $L \ge N+1$;
    \item \textbf{Property 2 (convexity):} 
    The point $\bm{b}^m$ 
is ``surrounded in all directions''
by the $\mathcal{B}_L(\bm{b}^m)$ neighborhood.
In mathematical terms,
$\bm{b}^m$ can be expressed as a convex combination
\begin{subequations}
\label{Eqn:ConvexCombination}
\begin{eqnarray}
    \bm{b}^m &=& \sum_{l=1}^{L} w_l \>\bm{b}^{k^m_l}
\\ \hbox{with}\quad 0 & \le & w_1, \ldots, w_L
\\ \hbox{and} \quad 1 & = & \sum\limits_{l=1}^L w_l
\end{eqnarray}
\end{subequations}
Note that the weights do not need to be unique, for instance, for $L>N+1$.
    \item \textbf{Property 3 (non-degeneracy):} The $L$ neighbors span the $N$-dimensional space. 
\end{itemize}

Properties 2 and 3 imply that for all hyperplanes passing through $\bm{b}^m$ the $L$ neighbors can be found on each side of the hyperplane.
In other words, for every unit vector $\bm{e}$
\begin{equation}
   \underset{i=1,...,L}{\min} \left\{ \bm{e} \cdot (\bm{b}^{k^m_i}- \bm{b}^m) \right\} < 0 < \underset{i=1,...,L}{\max} \left\{ \bm{e} \cdot (\bm{b}^{k^m_i}- \bm{b}^m) \right\}.
   \label{eq:isotropy}
\end{equation}
holds.
In practice, property 2 can be reformulated as a linear programming problem and solved with the ``scipy'' Python package.

The global maximum $\bm{b}^m$ is defined as a local maximum with $L = M-1$, i.e., $\bm{b}^m$ is inside the data set and its $FM$ is greater than any other point in the data set.
\begin{table}
\centering
\caption{Number of neighbors ($L$) for the global and local maxima of the rotor data.\label{table:L neighbors}}
\setlength{\tabcolsep}{15mm}
\begin{tabular}{c c} 
 \hline
\hline
                              & $L$    \\ 
\hline
Global maximum                & 187  \\ 
\hline
\multirow{5}{*}{Local maxima} & 7    \\ 
                              & 12   \\ 
                              & 19   \\ 
                              & 21   \\ 
                              & 47   \\
\hline
 \hline
\end{tabular}
\end{table}
Table~\ref{table:L neighbors} shows the number of neighbors $L$ for the global and local maxima neighbors, i.e., the size of their neighborhood.
As defined above, the global maximum has $M - 1 = 187$ nearest neighbors.
The $L$ values of the local maxima are all bigger than $N + 1 = 5$ corresponding to the definition.
The size of the local maxima neighborhood ranges from 7 to 47.

\begin{figure}
\centering
\includegraphics[width=1\linewidth]{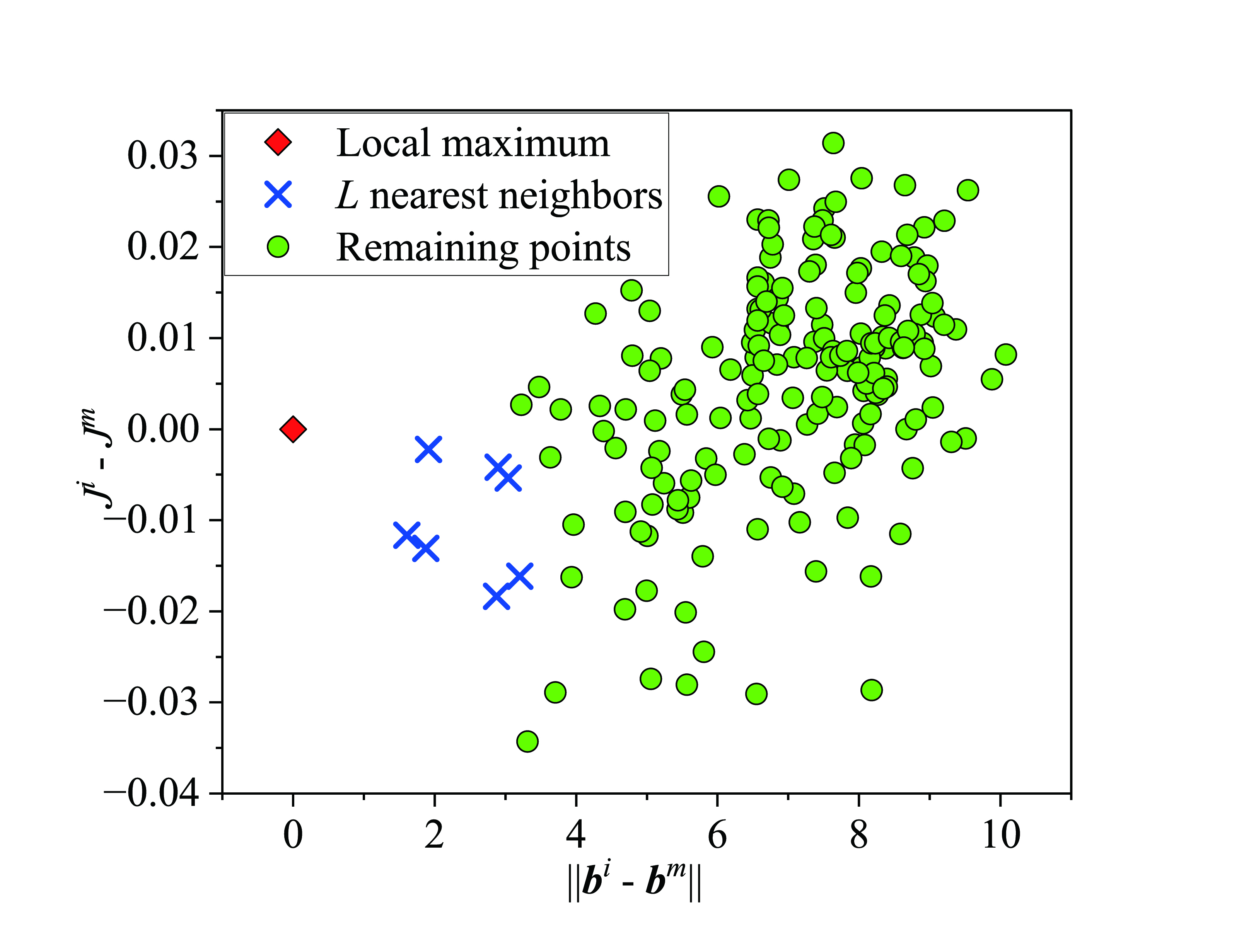}
\caption{\label{fig:local}
$FM$ discrepancy versus distance between a local maximum $\bm{b}^m$ (red diamonds), its $L$ neighbors (blue crosses) and the remaining data points (green dots).
The ID in the data set of this local maximum is 26.}

\end{figure}

Fig.~\ref{fig:local} depicts one example of a local maximum and its 7 neighbors.
The $FM$ of point (ID = 26) is higher than the $FM$ of its neighboring points, however other points in the data set have a higher $FM$.
There are 5 local maxima in the rotor data set, their IDs are 26, 52, 54, 64, and 113 and their respective $FM$s are 0.554, 0.567, 0.580, 0.570, and 0.582.
\begin{figure}
\centering
\includegraphics[width=1\linewidth]{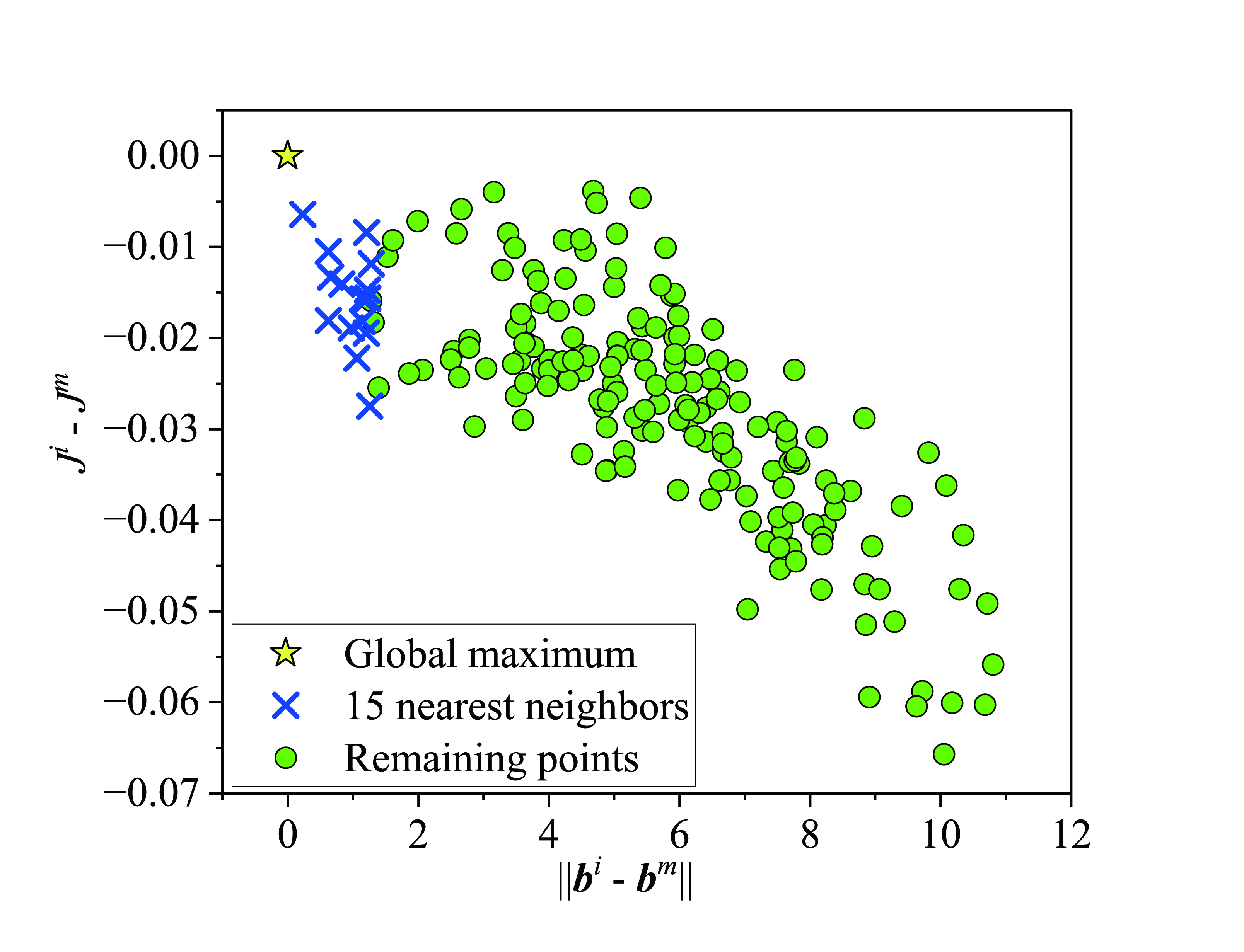}
\caption{\label{fig:global} 
$FM$ discrepancy versus distance between the global maximum $\bm{b}^m$ (ID = 170, $J^m=0.586$, yellow star), its 15 neighbors (blue crosses) and the remaining data points (green dots).
}
\end{figure}
Fig.~\ref{fig:global} displays the global maximum of the data and its 15 neighbors.
The global maximum has the highest $FM$, not only than its neighbors but also than all the other points in the data set.

If a point $\bm{b}^m$ does not satisfy property 2 then there is at least one direction where the maximality cannot be verified; 
Such point is referred in the following as ``one-sided'' maximum, see Fig.~\ref{fig:ConditionC}(b).
A typical example of one-sided maximum is a point at the border of the data set,
where the maximality in the direction pointing outside the data cannot be established.
As the local and global maxima, the one-sided maxima give information on the topology of the search space.
The detected one-sided maxima may also be at the frontier of isolated clusters of data.

 \begin{figure}
\centering
\includegraphics[width=1\linewidth]{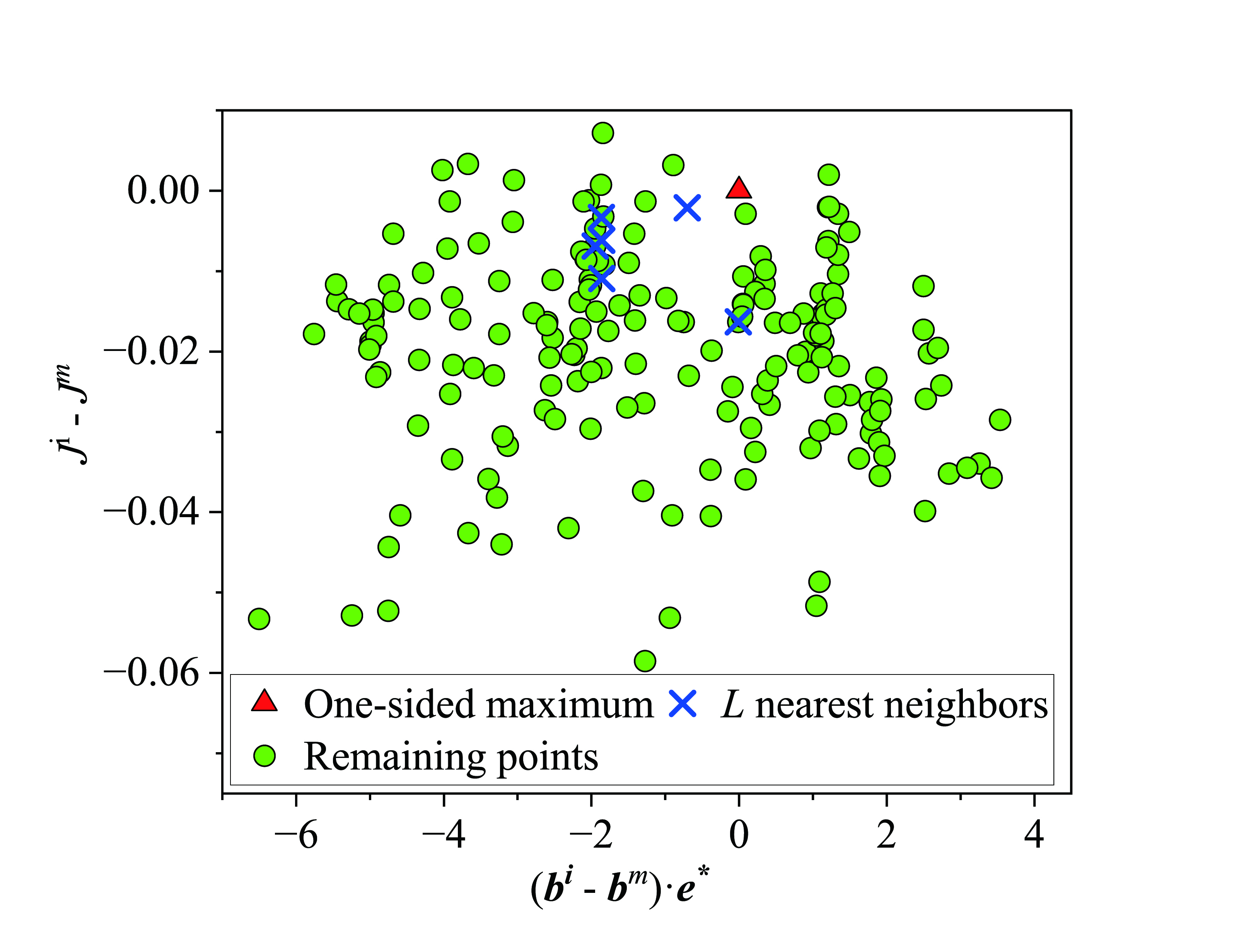}
\caption{\label{fig:One-sided minimum} 
 One-sided  local maximum $\bm{b}^m$ (red triangle), its $L$ neighbors (blue crosses) and the remaining data points (green dots).
 The data is projected following the vector $\bm{e}^*$, i.e., the direction where the neighbors are located only on one side.
}
\end{figure}
Fig.~\ref{fig:One-sided minimum} shows an example of a one-sided maximum for the rotor data.
The data points are projected on the direction $\bm{e}^*$ where the maximality of the point cannot be established.
Indeed, all the neighbors of the point $\bm{b}^m$ are located on one side---the negative side---of direction $\bm{e}^*$.
In this case, $\bm{b}^m$ is a local maximum for the negative side, but the absence of data on the positive side does not allow us to conclude that $\bm{b}^m$ is indeed a maximum for the positive side.
Vector $\bm{e}^*$ defines a  hyperplane passing through $\bm{b}^m$ and isolates the data on one side of the space.
Such a hyperplane is a linear classifier or separatrix that can be learned with Support Vector Machine (SVM).
In the SVM framework, $\bm{b}^m$ is labelled $-1$, all the others points are labeled 1 and we look for the linear classifier that separates the data at $-1$.

%

\begin{figure}
\centering
\includegraphics[width=1\linewidth]{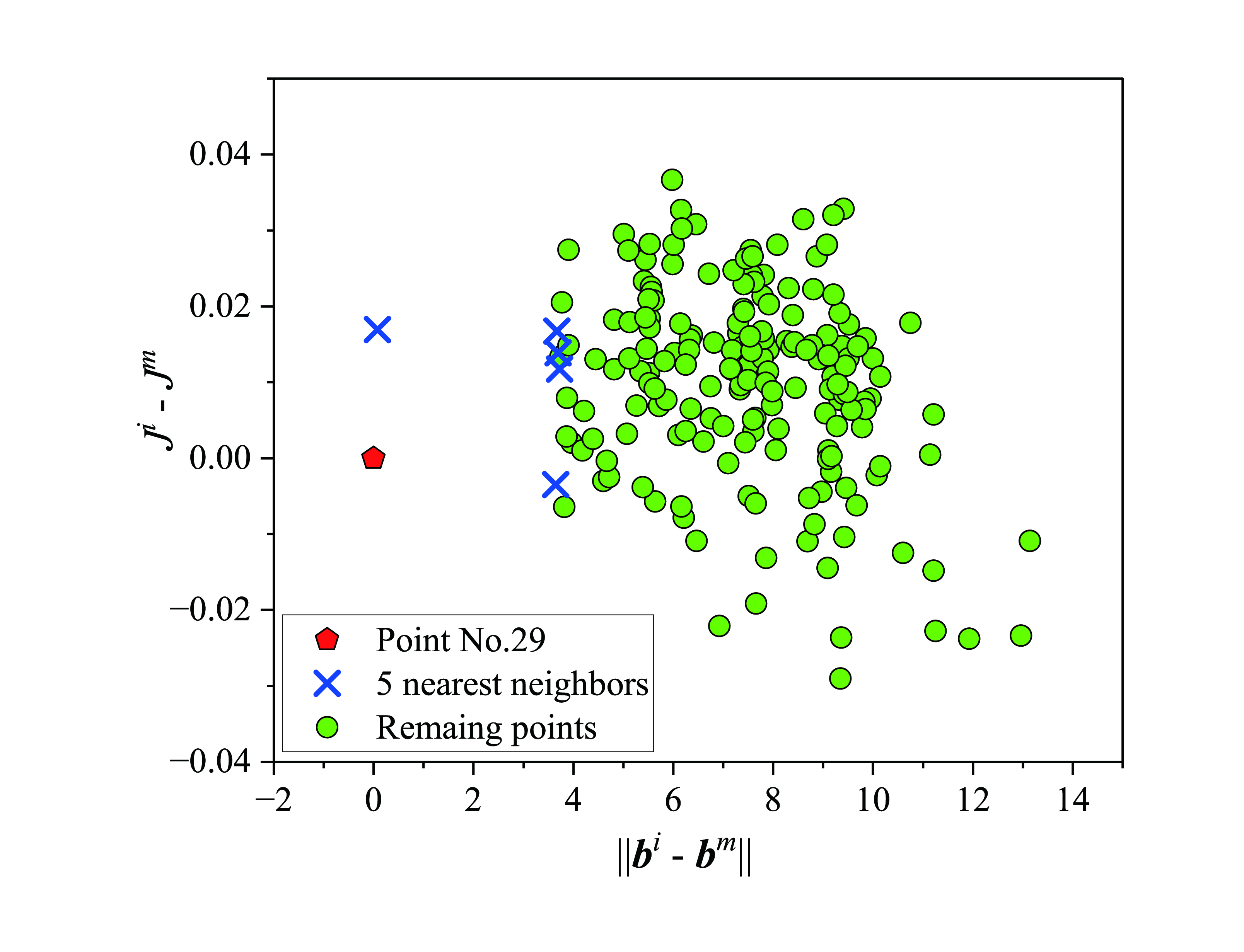}
\caption{\label{fig:29} 
$FM$ discrepancy versus distance between the point $\bm{b}^m$ (ID = 29, red pentagon), its 5 neighbors (blue crosses) and the remaining data points (green dots).
}
\end{figure}
Finally, the data is perturbed with high noise.
Fig.~\ref{fig:29} shows that despite being close to each other, the point (ID = 29) and its closest neighbor have a large $FM$ difference $\Delta J = 0.017$, which corresponds to $25.77\%$  of the difference between the maximum and the minimum $FM$.
The mean noise level, computed over the points closer than $0.5\%D$, is $N_L=0.007$, i.e., $10.61\%$ of the $FM$ range.

Such a discrepancy between two close points is expected to artificially introduce local or global maxima in the data and thus giving a false representation of the search space topology.
 
 The neighborhood analysis in this section indicates that the data set is noisy and presents maxima of different nature.
 The maxima may be a result of noise and the lack of data in the neighborhood of some points.
 In the following, we propose a method to curate such data sets.

\section{\label{sec:TAO}Topologically assisted optimization }
Data noise artificially introduces local maxima in the data that are detrimental for any gradient-based optimization method.
However, for reasonable levels of noise, the topology of the search space is preserved,
i.e., the plateau, valleys, hills and pits still shape the search landscape.
In particular, the global maximum remains in the same region.
We propose a data smoothing method to flatten the artificial maxima.
In this section, the topologically assisted optimization (TAO) for curating noisy data is described.
The method is based on an elastic response model (ERM, Section~\ref{sec:ERM}) to model the interaction between the data points.
TAO is then illustrated on a one-dimensional test function (Section~\ref{sec:1D test}).
Finally, the persistency of the rotor data is described (Section~\ref{sec:Persistency}) and the TAO-smoothed data is analyzed (Section~\ref{sec:TAO_Analysis}).

\subsection{\label{sec:ERM}Elastic response model}
TAO data smoothing is based on the perturbation of the data set with additive "anti-noise".
Let $\{\bm{b}^m,j^m\}_m^M$ be the "$\epsilon$-perturbed data set" 
where $j^m$ is the perturbed $FM$ of $\bm{b}^m$ such as
\begin{equation}
    |j^m - J^m| \leq \epsilon.
    \label{Eq:j_epsilon}
\end{equation}
The $\epsilon$ parameter is the maximum amplitude of the anti-noise introduced in the data.
As heuristic, the perturbation leading to a smoother data set follows an iterative elastic response model (ERM).
We note $(j_n^m)_n$ the sequence of perturbed data that follows the ERM.
The process is initialized with no initial perturbation, i.e., the original data:
\begin{equation}
    j^m_0  = J^m.
\end{equation}
For $n\geq 0$ each perturbed data point is corrected by the $FM$ difference with its $N+1$ neighbors and weighted by the inverse of their distance:
\begin{align}
\tilde{j}^m_{n+1} & = j^m_n + \cfrac{\alpha}{N+1} \sum^{N+1}_{i=1} \cfrac{j^{k^m_i}_n - j^m_n}{||\bm{b}^{k^m_i}-\bm{b}^m||} \nonumber \\
j^m_{n+1} & = \left\lbrace
\begin{array}{lcr}
     J^m -\epsilon & \text{if} & \tilde{j}^m_{n+1} \leq J^m-\epsilon  \\
     \tilde{j}^m_{n+1} &  \text{otherwise} & \\
     J^m +\epsilon & \text{if} & \tilde{j}^m_{n+1} \geq J^m+\epsilon
\end{array} \right. \label{Eq:ERM}
\end{align}
Note that the perturbation is bounded to the limit set in equation~\eqref{Eq:j_epsilon}.
The coefficient $\alpha$ sets the correction rate, it is chosen to balance smooth correction and quick convergence.
The procedure is iterated until the difference between two consecutive steps is below a given threshold $\beta$:
\begin{equation}
\sum_{i=1}^M |j_{n}^i - j_{n-1}^i| \leq \beta.
\end{equation}

The smoothing is then parametrized by the maximum anti-noise level $\epsilon$.
For $\epsilon = 0$, no anti-noise is introduced and the data remains the same.
As $\epsilon$ increases, the anti-noise smooths the irregularities of increasing size, reducing progressively the number of maxima.
For $\epsilon = \cfrac{1}{2}(\underset{m}{\max}\; J^m - \underset{m}{\min}\; J^n)$,
all the maxima are smoothed out as the anti-noise can theoretically bring the maximum and minimum $FM$ at the same level.
The ``persistence curve'' is defined as the evolution of number of maxima $N_{\rm max}$ as a function of $\epsilon$:
\begin{equation}
    \epsilon \mapsto N_{\rm max}.
\end{equation}
The persistence curve gives the stability of the maxima with increasing noise \citep{Kasten2016aom} and allows the distinction between the maxima intrinsic to the search space and the maxima artificially introduced by the noise related maxima.
Thus, the last maximum to persist to increasing $\epsilon$ is the global maximum.

\subsection{\label{sec:1D test}Qualitative scenarii exemplified for a 1D example}
In this section, we illustrate the TAO methodology on a one-dimensional function including many maxima:
\begin{equation}
    J(b) = e^{-b^{2} } -\frac{\left ( b+15 \right )}{40}  \sin \left ( 2\pi  b\right ).
    \label{Eq:1DJ}
\end{equation}
\begin{figure}
\centering
\includegraphics[width=0.5\textwidth]{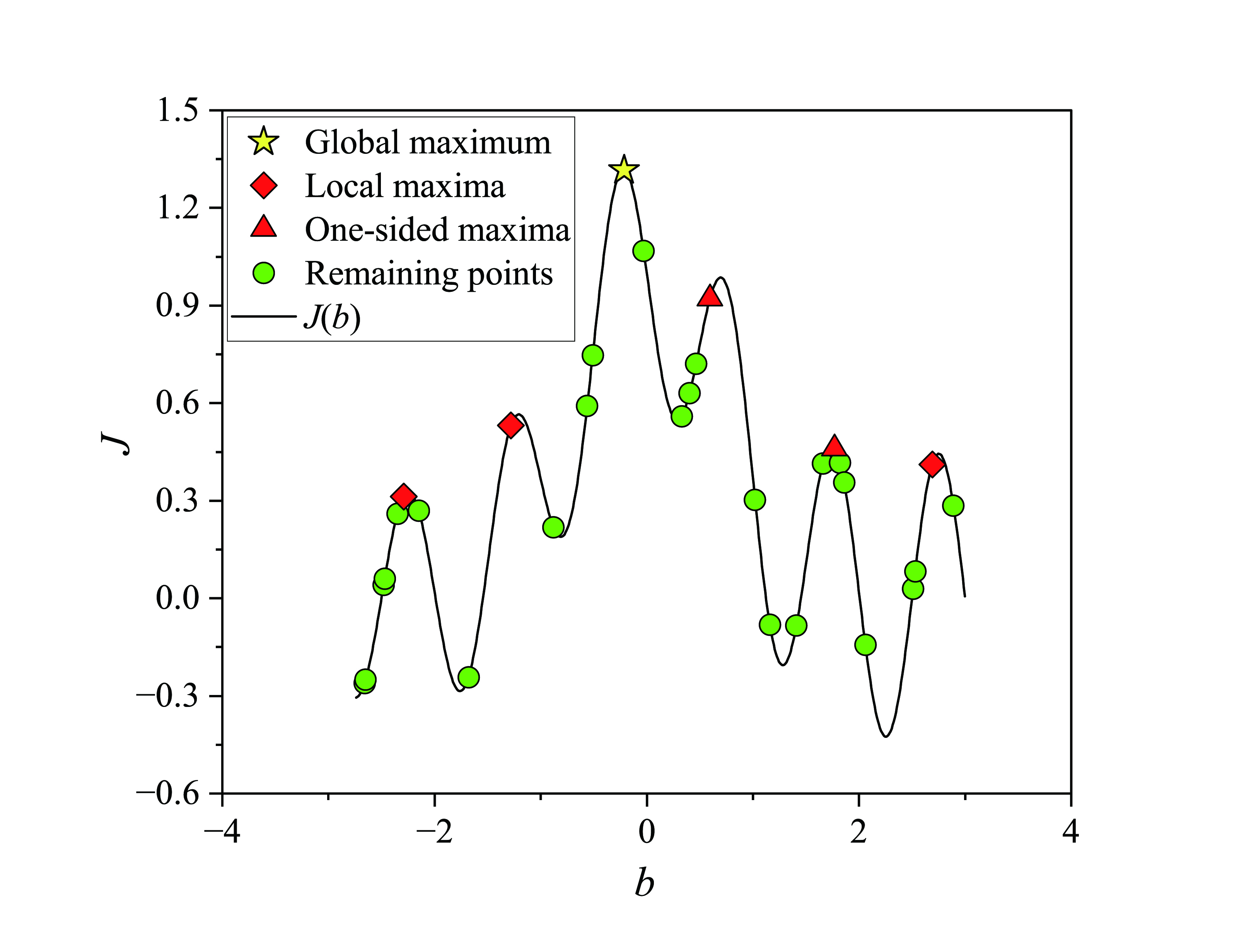}
\caption{\label{fig:1D} Depiction of the 30 random data points for the one-dimensional test function~\eqref{Eq:1DJ}.
}
\end{figure}
The function $J$ includes two terms: one exponential term $e^{-b^2}$ that defines a smooth shape with a unique maximum at $b=0$,
and a sinusoidal term that acts as source of noise.
The parameters of the sinusoidal function are chosen such as $J$ includes 6 maxima between -3 and 3, see Fig.~\ref{fig:1D}.
The data set is built by sampling 30 random points in that range.
The data, displayed in Fig.~\ref{fig:1D}, includes 6 maxima: 2 one-sided maxima, 3 local maxima and the global maximum located at $b=-0.22$.
The difference between the minimum and maximum $FM$ is $1.80$.
\begin{figure}
\centering
\includegraphics[width=0.35\textwidth]{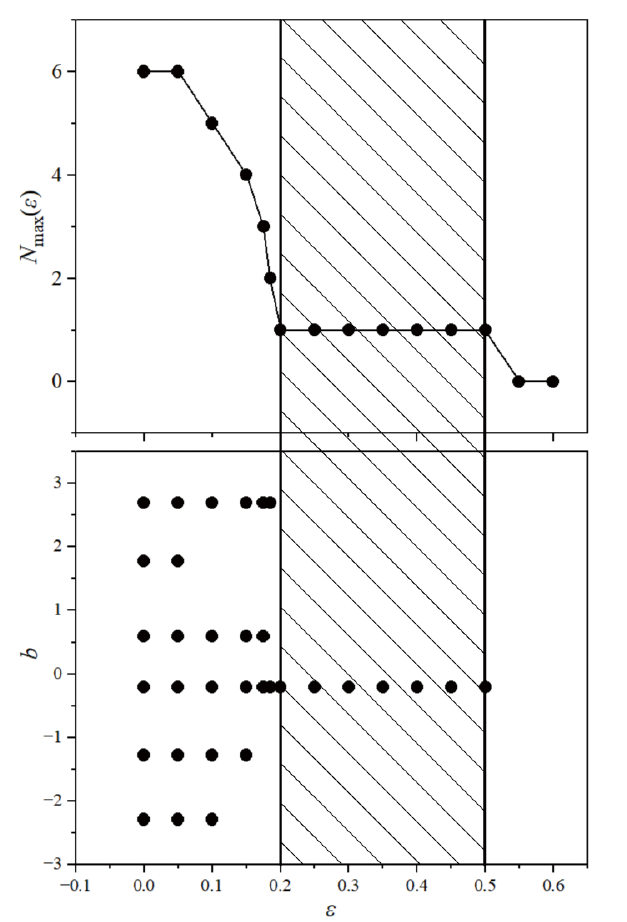}
\caption{\label{fig:Persistent1D} Persistence curve (top) and persistence maxima (bottom) for the one-dimensional example. The dashed region shows where the global maximum is the only maximum of the data.}
\end{figure}

Fig.~\ref{fig:Persistent1D} shows the persistence curve and the persistence maxima for the one-dimensional data.
Expectedly, as $\epsilon$ increases, the number of maxima decreases.
The last maximum to remain is the global maximum at $b=-0.22$.
The number of maximum reaches $N_{\rm max}=0$ from $\epsilon=0.55$.

\begin{figure}
\centering
\includegraphics[width=0.5\textwidth]{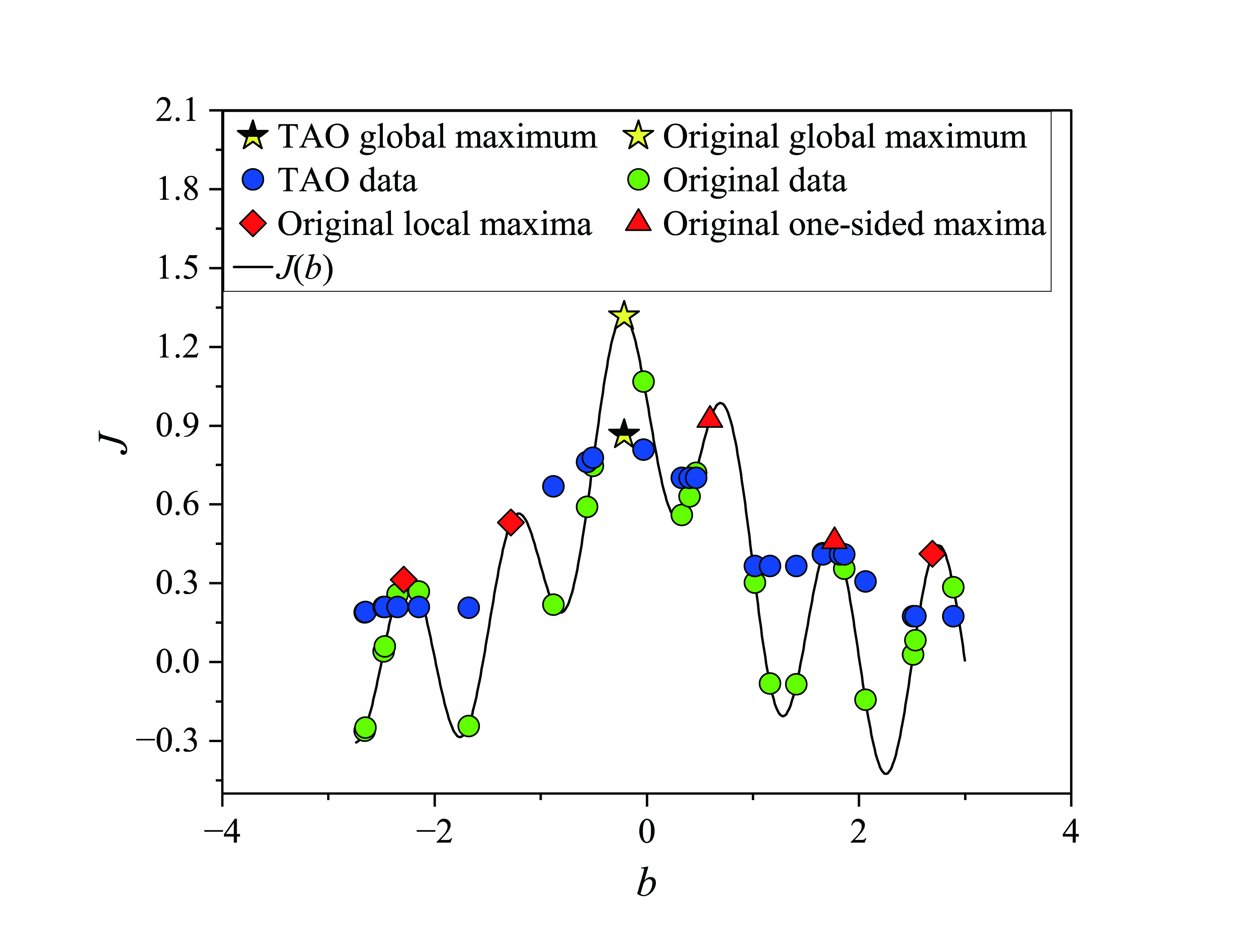}
\caption{\label{fig:1D_1max} Depiction of $\epsilon$-perturbed data set for the one-dimensional example.
} 
\end{figure}
Fig.~\ref{fig:1D_1max} shows the $\epsilon$-perturbed data set for $\epsilon=0.45$, i.e., when only the global maximum remains.
The last maximum is the global maximum of the data.
For the remaining points, we note that the data has been flattened at some regions, smoothing out the maxima.

This one-dimensional example demonstrate TAO's ability to flatten the artificial maxima due to noise and preserves the global maximum of the data.

\subsection{\label{sec:Persistency}$\epsilon$ persistency of the rotor data}
\begin{figure}
\centering
\includegraphics[width=0.35\textwidth]{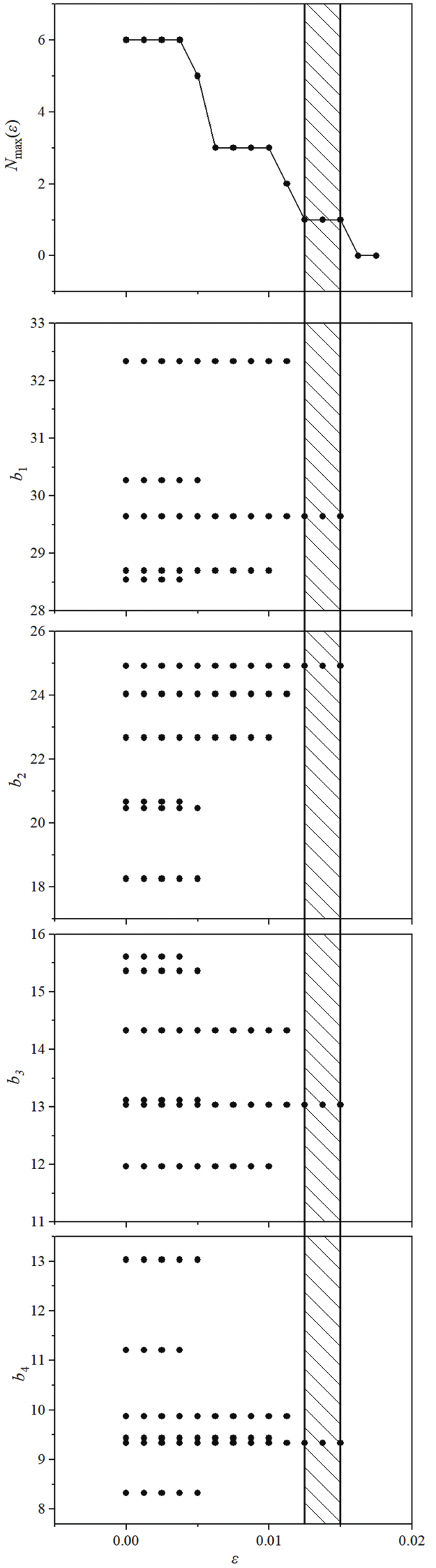}
\caption{\label{fig:Persistency} Persistence curve (top) 
 and $\epsilon$ persistent maxima (others) for the rotor data. The dashed region represents the range where the global maximum is the only maximum of the data.}

\end{figure}
Now, we analyze the rotor data with TAO.
Fig.~\ref{fig:Persistency} shows the persistence curve and the components of the persistent maximum points $\epsilon$.
The rotor data has initially 24 maxima: 18 one-sided maxima, 5 local maxima and the global maximum ($\bm{b}^*=[29.64, 24.91, 13.04, 9.33]^{\intercal}$).
As expected, the number of maxima decreases as $\epsilon$ increases.
An anti-noise level of $\epsilon=0.012$ smooths out all maxima expect the global one.
$\epsilon=0.012$ represents approximately $18\%$ of the difference between the maximum and minimum of the data,
which corresponds to $1.7$ times the estimated noise level $N_L$ in Section\ref{sec:Neighbor analysis}.
The global maximum remains beyond this noise cancellation until $\epsilon=0.015$,
which is close to $23\%$ of the $FM$ range in the data and 2.4$N_L$.

The persistence curve show that anti-noise efficiently smooths out the artificial local maxima but also reveals the persistence of the global maximum to large levels of noise.

\subsection{\label{sec:TAO_Analysis} TAO analysis}
In this section, we describe the search space for the rotor data before and after smoothing with TAO.

\begin{figure}
\centering
\includegraphics[width=1\linewidth]{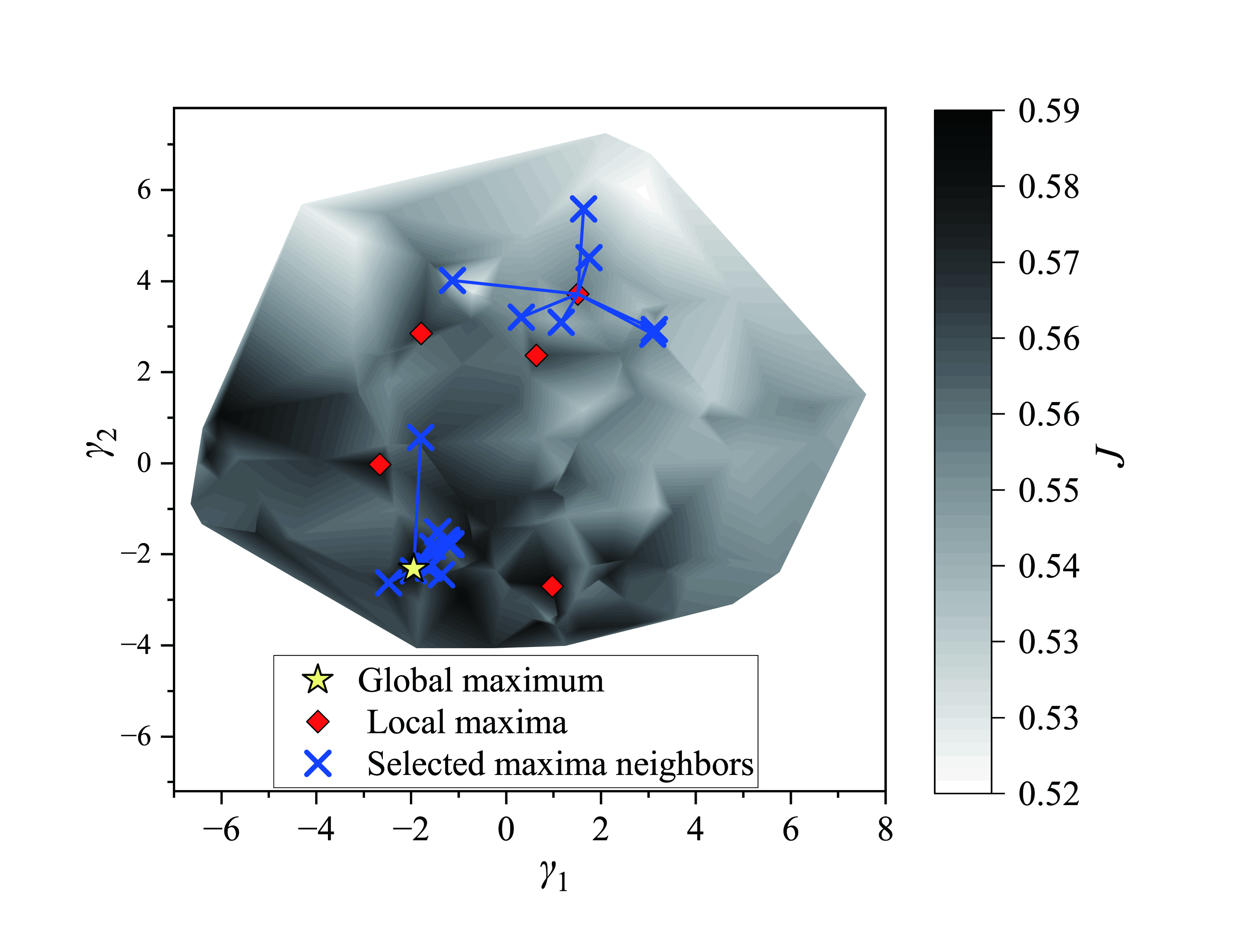}
\caption{\label{fig:g_l_min} Proximity map of the rotor data.
Dark regions denote poor performances and light regions good performances.
}
\end{figure}
Fig.~\ref{fig:g_l_min} displays the rotor data, the global maximum and local maxima on a proximity map, i.e., a 2D representation of the search space.
The proximity map is based on classical multi-dimensional scaling (MDS) and is a projection of the data on a two-dimensional space that preserves the distance between points~\citep{Kaiser2017ifac,LiA2022jfm}.

MDS is a dimensional reduction method that consists of extracting the two main features of the flow ($\gamma_1$ and $\gamma_2$) by applying a proper orthogonal decomposition on the distance matrix of the feature vector $\bm a$.
The vectors $\gamma_1$ and $\gamma_2$ spawn a two-dimensional space where all the data is projected;
It is the optimal projection, regarding the $L^2$-norm, that preserves the distances between the states.
In Fig.~\ref{fig:g_l_min}, the good-performing regions (black), i.e., high $FM$ regions is located around the bottom of the map, and the top of the map denotes to the bad-performing region (white), i.e., low $FM$ regions. 
As a result, we see that there is a global gradient from top to bottom.
The global maximum (yellow star) is located close to the boundary of the data region;
It is surrounded by its neighbors (blue crosses).

Five local maxima are shown in Fig.~\ref{fig:g_l_min} as red diamonds.
The point with ID = 26 is selected as an example to show that the neighbors of the local maximum surround the maximum.

\begin{figure}
\centering
\includegraphics[width=1\linewidth]{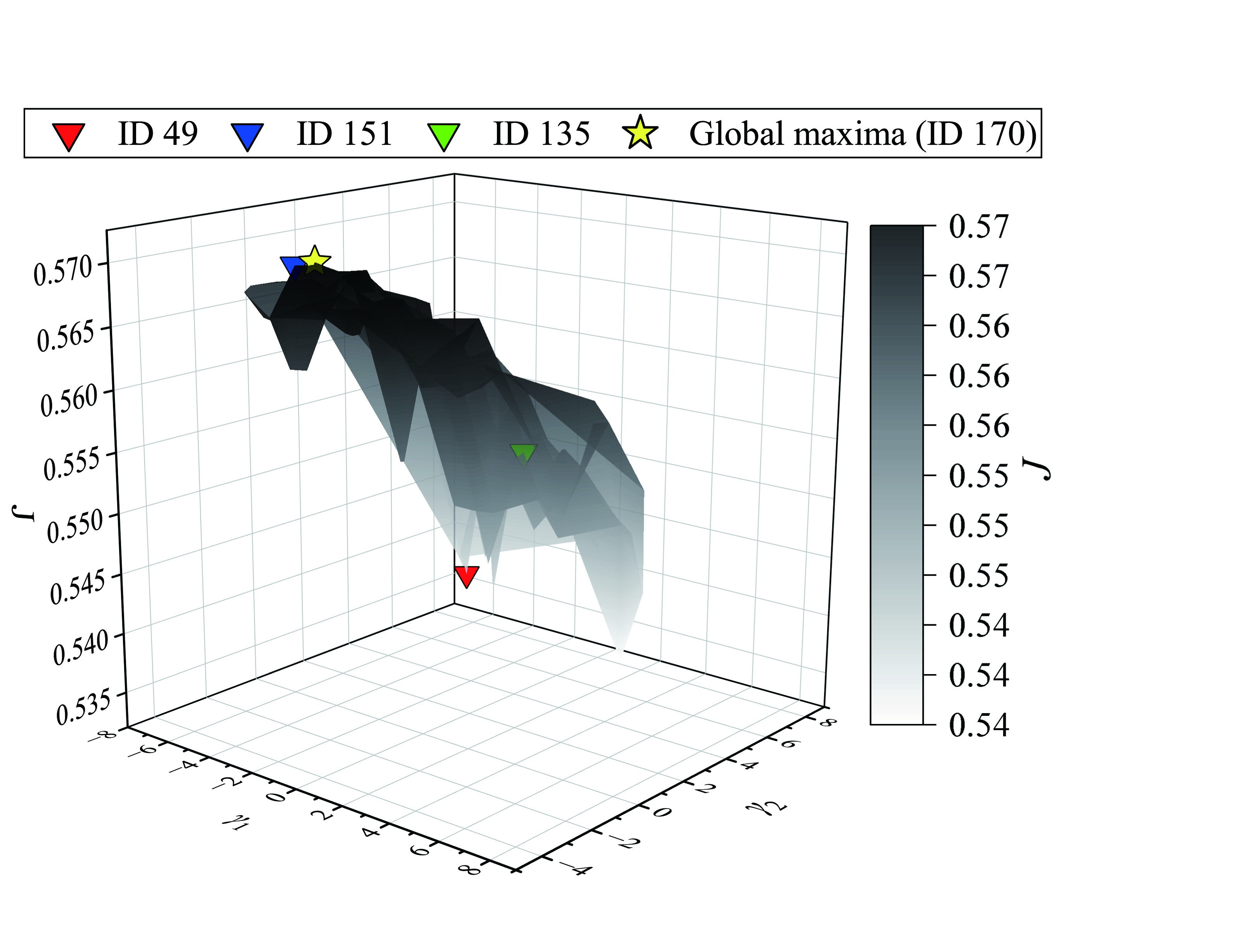}
\caption{\label{fig:single global maximum} Proximity map of the rotor data after TAO-smoothing with an anti-noise level of $\epsilon=0.015$.}
\end{figure}

From Fig.~\ref{fig:Persistency}, $N_{max}$ equals 1 when $\epsilon$ ranges from 0.012 to 0.015.
A random value $\epsilon$, 0.015, is chosen from this range to show the smoothing result of TAO.
Fig.~\ref{fig:single global maximum} shows that there is only a single global maximum in the data region, which means that the data region is smoothed.
As a result, TAO is verified in high-dimensional problems.

From Fig.~\ref{fig:single global maximum}, the control landscape presents a steep slope going from the global maximum towards a point (ID = 49).
 
There is almost a plateau from the blue triangle (ID = 151) to the yellow star (global maximum, ID = 170), while a steep slope starting from the yellow star to the green triangle (ID = 135) then to the red triangle (ID = 49).
As a result, there is no quadratic response model that can capture such a sudden cliff.
To clearly depict it, the data is projected along the vector $\bm{e}^H$ that is orthogonal to the hyperplane passing through the global maximum $\bm{b}^*$ and minimizing the number of points on one side.
For points on the border of the domain, such hyperplane defines the local frontier of the data.
In practice, the vector $\bm{e}^H$ is derived by downhill simplex minimizing the number of positive values of $(\bm{b}^m - \bm{b}^*)\cdot \bm{e}^H$.
The value of $\bm{e}^H$ is:

\begin{equation}
    \bm{e}^H = \begin{bmatrix}-0.19
 \\   0.96
 \\0.01
 \\-0.10
\end{bmatrix} .    
    \label{Eq:barrier vector}
 \end{equation}

\begin{figure}
\centering
\includegraphics[width=0.5\textwidth]{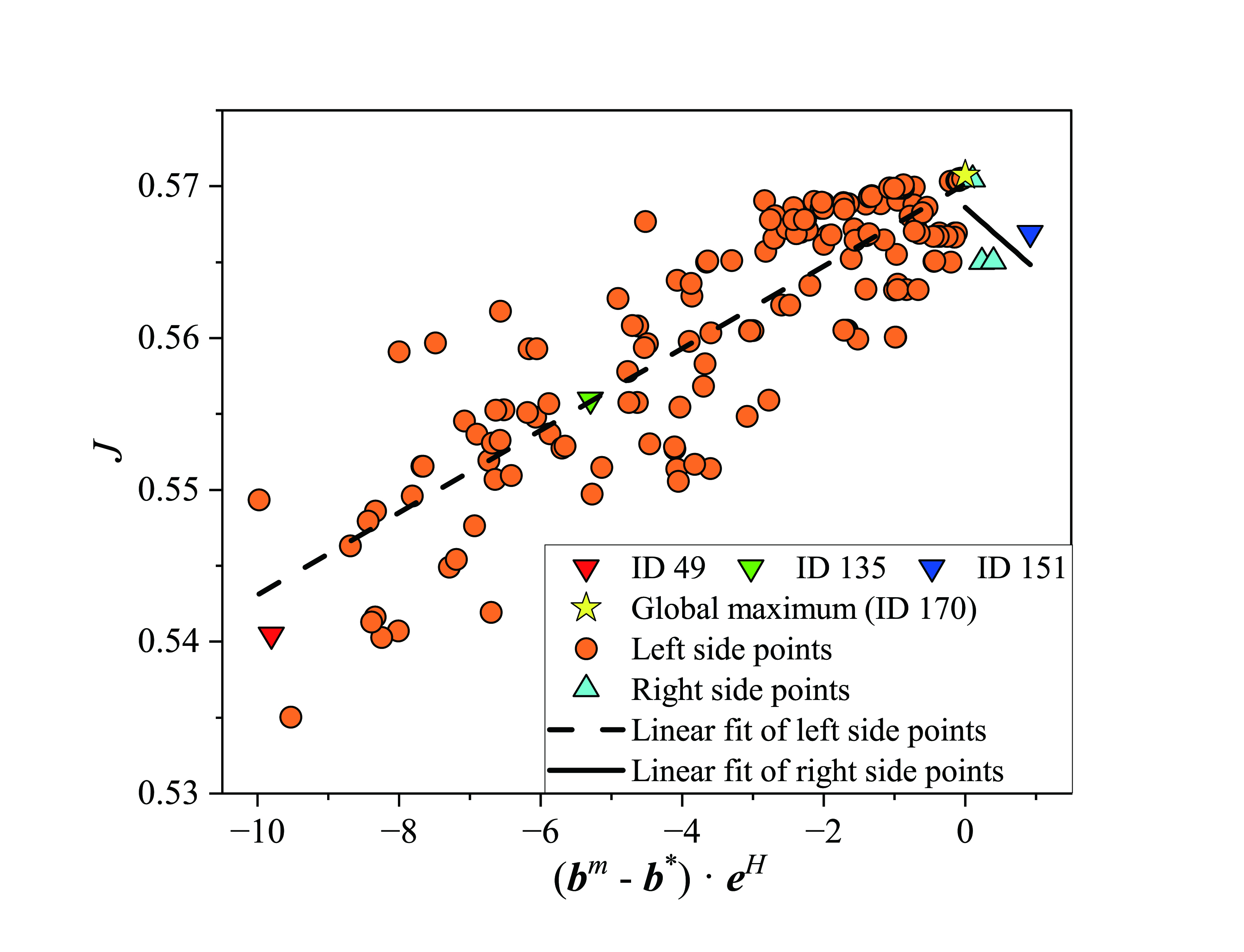}
\caption{\label{fig:TAO_bias} TAO data projected along the $\bm{e}^H$.
} 
\end{figure}
Fig.~\ref{fig:TAO_bias} depicts the data projected along the direction $\bm{e}^H$.
First, the projection shows, independently of the proximity map, that the global maximum of the data is located inside the data,
i.e., not on the boundary.
Second, the projection also reveals the cliff in the data and the sharp descent.
The same marks of inverted triangles and yellow star in Fig.~\ref{fig:TAO_bias} and Fig.~\ref{fig:single global maximum} clearly show the cliff and the sharp descent. 

The distribution of the data is reminiscent of an incipient separation on an airfoil.

\begin{figure}
\includegraphics[width=0.49\textwidth]{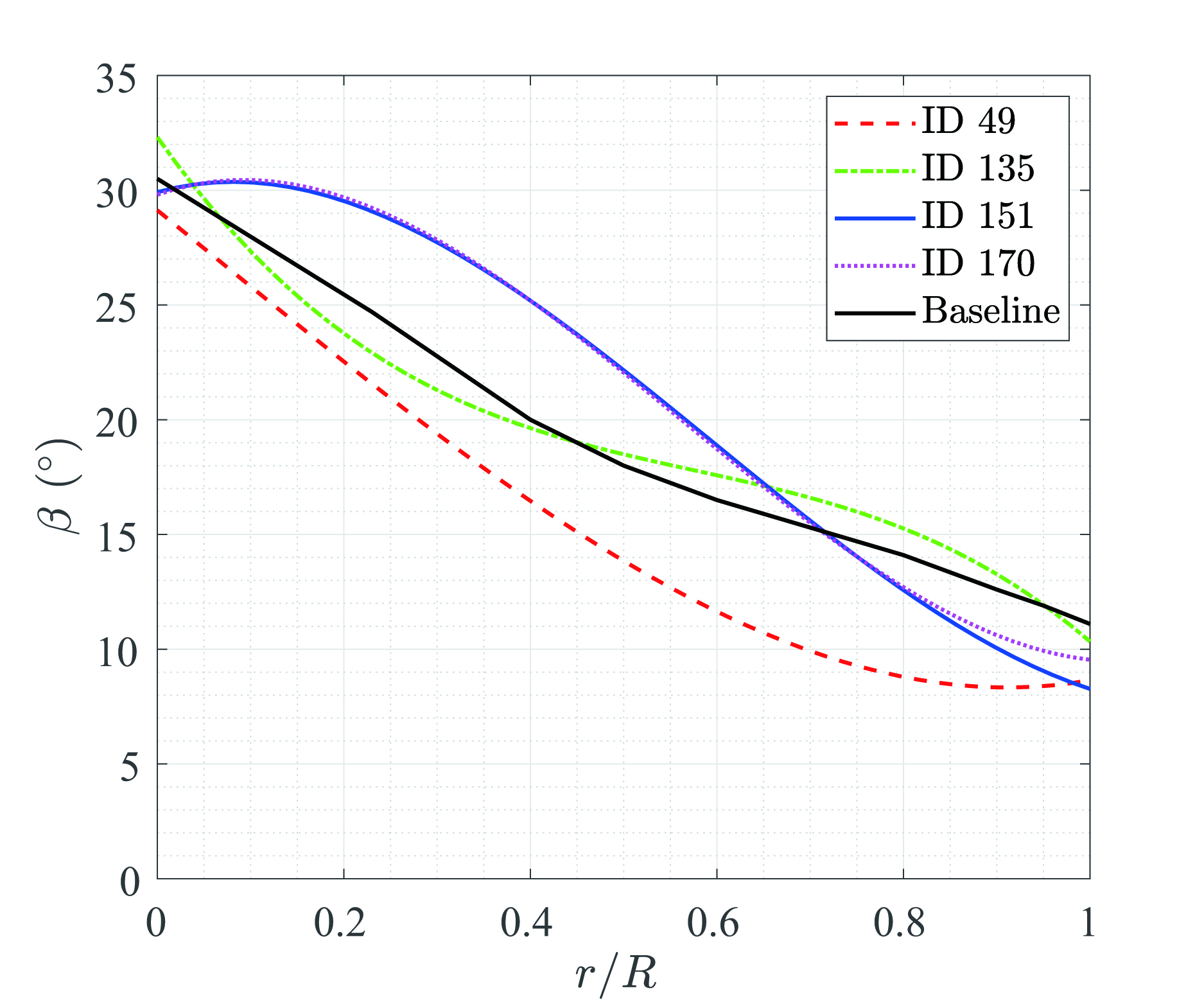}
\caption{Twist angle distribution in the radial direction for individual number of 49, 135, 151, and 170. }
\label{fig:four_points}
\end{figure}

\begin{figure}
\includegraphics[width=0.49\textwidth]{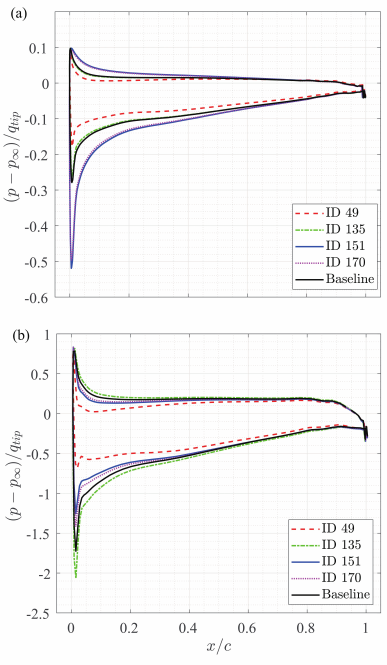}
\caption{Pressure coefficient distribution on the blade cross section. (a) Radial position of $r/R = 0.3$;(b) Radial position of $r/R = 0.9$. $q_{tip}$ is the dynamic pressure based on the blade tip velocity. }
\label{fig:four_points_cp}
\end{figure}

Fig.~\ref{fig:four_points} presents the geometry of four typical configurations, which includes the optimal design (ID 170), the leftmost point (ID 49) and the rightmost point (ID 151), and a point in the middle (ID 135) of Fig.~\ref{fig:single global maximum}.
The optimum point (ID 170) shows higher twist angle in the inner region ($r/R < 0.7$) and lower angle in the outer range ($r/R > 0.7$) with respect to the baseline design. The individual design with ID 49 has lower twist angle in all radial positions. The rightmost point in Fig.~\ref{fig:single global maximum} is the individual ID 151, and it has similar twist angle distribution as the baseline model, while lower twist angles are observed near the blade tip. 


The pressure distributions for the four typical designs are shown in Fig.~\ref{fig:four_points_cp}. Two radial positions, i.e., $r/R = 0.3$ and 0.9, are selected to represent the inner and outer ranges of the blade. In the inner range, the two cases that have a lower geometric twist angle (ID 49 and 135) with respect to the baseline model show a lower pressure difference between the suction and pressure sides. On the contrary, the two cases which have higher geometric twist angle (ID 151 and 170) with respect to the baseline model show higher pressure difference. The consistency between the twist angle and pressure difference means that the twist angle is proportional to the aerodynamic angle of attack of the blade cross section in this research.

In the outer range, the observation is similar to that in the inner range. The case ID 135 which has the highest twist angle corresponds to the largest pressure difference.  For the three cases that have a lower twist angle than the baseline model (ID 170, 151, and 49), the pressure difference decreases as the twist angle decreases. 

 The $x$ and $z$ forces (axes defined in Fig.~\ref{fig:chord_pitch}) on the blade cross section $r/R = 0.3$ are used to calculate the local thrust, power, and power loading, which are shown in Table ~\ref{table:cross_section0d3}. The local thrust is the force in the $z$ direction. The local power is calculated by the local thrust in $x$ direction multiplied by the radial distance to the hub center and rotation speed. The local power loading is an indicator of the local aerodynamic performance, which is analogous to the local lift-to-drag ratio of an airfoil. It can be seen that the optimal design (ID 170) and the right most point in Fig.~\ref{fig:TAO_bias} (ID 151) have lower local power loading in comparison to the other three models. By the same method, the local thrust, power, and power loading at $r/R = 0.9$ are shown in Table ~\ref{table:cross_section0d9}. The local power loading of the optimal design (ID 170) and the right most point in Fig.~\ref{fig:TAO_bias} (ID 151) are the two best designs. It can be concluded the aerodynamic performance of the cross section in the outer range is the leading factor here, as the dynamic pressure in the outer range is higher.

\begin{table}
\centering
\caption{Local thrust and power in the cross section $r/R = 0.3$.
\label{table:cross_section0d3}}
\setlength{\tabcolsep}{2mm}
\begin{tabular}{c c c c c c} 
\hline
\hline
 & ID 49 & 135 & 151 & 170 & Baseline     \\ 
\hline
${dT}/{dr}$ (N/m) &4.47      & 5.58   & 6.91   & 6.50  & 5.64   \\  %
\hline
${dP}/{dr}$ (W/m) & 28.62    & 36.84  & 54.06  & 51.93 & 38.95\\ 
\hline
Power loading & 0.1560 & 0.1514 & 0.1278 &0.1252 & 0.1449\\ 
\hline
\hline
\end{tabular}
\end{table}

\begin{table}
\centering
\caption{Local thrust and power in the cross section $r/R = 0.9$.
\label{table:cross_section0d9}}
\setlength{\tabcolsep}{2mm}
\begin{tabular}{c c c c c c} 
\hline
\hline
 & ID 49 & 135 & 151 & 170 & Baseline     \\ 
\hline
${dT}/{dr}$ &16.66 & 24.80 & 21.00 &21.5271  &22.9442   \\  %
\hline
${dP}/{dr}$ & 156.55& 266.94 & 194.36& 203.42 & 242.69 \\ 
\hline
Power loading & 0.1064 &  0.0929& 0.1081  &0.1058  & 0.0945\\ 
\hline
\hline
\end{tabular}
\end{table}

\section{Conclusions and outlook}
\label{sec:Conclusion}

We conduct a numerical rotor blade optimization at the hovering condition 
for maximum figure of merit ($FM$) based on the Reynolds-averaged Navier-Stokes equation
at a Reynolds number of $5.6 \times 10^4$ (based on the chord length and tangential velocity at the blade tip).
More specifically, 
the twist angle distribution of a quadcopter rotor is optimized 
using a genetic algorithm (NSGA-II). 
The objective function is the figure of merit
under fixed rotation speed of 6000 $RPM$. 
Evaluating a single shape is computationally prohibitively expensive:
A converged ANSYS CFX simulation requires 
about 25000 iterations corresponding to about 2600 core hours at a quiescent condition.
Hence, the NSGA-II optimization is performed with pre-converged data after 1000 iterations to keep the computational load manageable.
This approach comes with significant uncertainty of the objective function values.
The optimization is deemed converged 
after 4 generations with 50 individuals,
leading to 200 simulations in total.

A data neighborhood analysis reveals that small changes
of the optimization parameters can lead to large changes in the objective function.
For instance, a parameter variation of 0.5\% of the data range
may lead to an non-physical objective function variation of 25.77\%. 
This noise leads also to non-physical gradients and  spurious local maxima.
A novel topologically data analysis
aims to assess and smooth out the corresponding noise of the objective function.
Starting point is the definition of extrema
following discrete scalar-field topology \cite{Kasten2016am}.
This definition avoids highly biased smoothing filters 
and parametric differentiation which cannot be performed under this noise.
We propose a $\epsilon$-bounded ``anti-noise'' 
to the objective function values which minimizes the number of maxima.
This anti-noise is constructed with a dynamic elastic response model.
The number of maxima decreases from 6 at $\epsilon=0$ to 1 at $\epsilon=0.012$
corresponding to 18.27\%  
of the data range for the objective function.
This value can be considered as a measure 
for uncertainty assuming monomodal objective function.

The global maximum is associated with a monotonic increase
of the twist angle with increasing radius, as expected.
The optimal parameters are inside the data region 
but not to far from the boundary.
The objective function decreases rapidly 
in the direction towards the data boundary
and slowly in the opposite direction.
The rapid decay can be associated with beginning separation
and the slow decay with gradual build up of lift.
The strong asymmetry explains the failing attempts
to model the objective function accurately by a second-order polynomial.

The combination of global optimizer 
and topological data analysis
is termed ``Topologically Assisted  Optimization (TAO)''.
This data analysis is a critical enabler
for optimization and surrogate modeling.
The data topology suggests the most fitting optimization approach.
For instance, a single optimum in the data
would encourage gradient-based ascends.
Few optima in the high-dimensional parameter space
might indicate random-restart gradient-based ascends.
Apparent noise and many optima would rule these approaches out
and encourage evolutionary approaches, like the chosen genetic algorithm.
If the optimum is located at the data boundary,
ridgeline extrapolation \cite{Fernex2020prf} is a method of choice
and promises significant performance improvements leaving the data region.
In contrast, an optimum inside the data indicates
that little performance improvement can be gained
in the vicinity of the maximum with small gradients.
The topology of the data also guides the development of a surrogate model.
We believe that future optimization algorithms 
will increasingly incorporate data topology for the mentioned reasons.

\begin{acknowledgments}
Bernd Noack acknowledges support
by the National Science Foundation of China (NSFC) through grants 12172109 and 12172111, 
by Guangdong province, China, 
via the Natural Science and Engineering grant  2022A1515011492 
and by the Shenzhen Research Foundation for Basic Research, China,  
through grant JCYJ20220531095605012.

Yannian Yang would like to thank AVIC General Aircraft Research Institute (Grant No. AG-EX$\_$HT$\_$024) for the support.

Angelo Iollo  acknowledges support from the European Union’s Horizon 2020 research and innovation programme under the Marie Skłodowska-Curie grant agreement No. 872442.
\end{acknowledgments}

\section*{Data Availability Statement}
The data that support the findings of this study are available from the corresponding author upon reasonable request.

\section*{References}
\bibliography{aipsamp}

\end{document}